\documentclass[reprint,aps,amsmath,amssymb,amsfonts,showpacs,nofootinbib]{revtex4-1}
\usepackage{graphicx}
\usepackage{bib_macro}
\usepackage{natbib}
\usepackage{placeins}
\usepackage[T1]{fontenc}
\usepackage{float}
\usepackage{epstopdf}
\usepackage[version=3]{mhchem}
\newcommand*{\fontpackage}{}
\renewcommand*{\fontpackage}{%
 txfonts%
}
\usepackage{color}
\usepackage{hyperref}

\hypersetup{
    colorlinks=true,
    linkcolor=red,
    citecolor=blue,
} 

\newcommand*{\fontpackageoptions}{}
\renewcommand*{\fontpackageoptions}{
 varg
}
\newcommand*{\isomathoptions}{}
\renewcommand*{\isomathoptions}{%
OMLmathsans,%
 sfdefault=zavm
}
\usepackage[\fontpackageoptions]{\fontpackage}
\usepackage[\isomathoptions]{isomath}
\usepackage{esint}
\newcommand{\lapl}{\upDelta}
\newcommand{\Lambdaup}{\mathrm{\Lambda}}
\renewcommand*{\dot}[1]{%
  \overset{\mbox{\large$.$}}{#1}}
\renewcommand{\rho}{\varrho}

\begin{document}
\title{Spherical collapse and halo mass function in $\mathbold{f(R)}$ theories}
\author{$^{1,2}$Michael Kopp}
\email{michael.kopp@physik.lmu.de}
\author{$^{2,3}$Stephen A. Appleby}
\email{stephen.appleby@ewha.ac.kr}
\author{$^{1,2,4}$Ixandra Achitouv}
\email{achitouv@usm.uni-muenchen.de}
\author{$^{1,2,5}$Jochen Weller} 
\email{jochen.weller@usm.lmu.de}
\affiliation{$^{1}$University Observatory, Ludwig-Maximillians University Munich,  \\ Scheinerstr. 1, 81679 Munich, Germany \\ 
$^{2}$Excellence Cluster Universe, Boltzmannstr. 2, 85748 Garching, Germany \\
$^{3}$Institute for the Early Universe WCU, Ewha University, \\ Seoul, Korea \\
$^{4}$Laboratoire Univers et Th\'eories (LUTh), UMR 8102 CNRS, Observatoire de Paris,
Universit\'e Paris Diderot, \\ 5 Place Jules Janssen, 92190 Meudon, France\\
$^{5}$Max-Planck-Institut f\"{u}r extraterrestrische Physik, \\ Giessenbachstrasse, 85748 Garching, Germany}

\begin{abstract}
We compute the critical density of collapse for spherically symmetric overdensities in a class of $f(R)$ modified gravity models. For the first time we evolve the  Einstein, scalar field and non-linear fluid equations, making the minimal simplifying assumptions that the metric potentials and scalar field remain quasi-static throughout the collapse. Initially evolving a top hat profile, we find that the density threshold for collapse depends significantly on the initial conditions imposed, specifically the choice of size and shape. By imposing `natural' initial conditions, we obtain a fitting function for the spherical collapse $\delta_{c}$ as a function of collapse redshift, mass of the overdensity and $f_{\rm R0}$, the background scalar field value at $z=0$. By extending $\delta_{c}$  into drifting and diffusing barrier within the context of excursion set theory, we obtain a realistic mass function that might be used to confront this class of scalar-tensor models with observations of dark matter halos. The proposed analytic formula for the halo mass function was tested against Monte Carlo random walks for a wide class of moving barriers and can therefore be applied to other modified gravity theories.
 \end{abstract} 
\maketitle 
\section{Introduction}
Einstein's theory of General Relativity (GR) \cite{E15} has withstood nearly one century of experimental testing. Many of its predictions have been confirmed through high precision laboratory and solar system experiments, and more recently with astrophyiscal and cosmological data (see \cite{W06} for a review). Testing GR on the largest scales is a field still in its infancy, as probing cosmological distances is technically challenging. However current and future surveys mapping the large scale structure of the universe already provide a powerful tool to constrain gravity models \cite{BOSS,DES,Pan,Euclid}. 

In spite of its success, there are theoretical reasons to believe that GR is not a fundamental theory of gravity. In particular one expects corrections to the Einstein-Hilbert action through 1-loop corrections induced by matter \cite{DFCB77}. In addition, there are puzzling cosmological observations such as the recently discovered accelerated expansion \cite{R98,P99,DETF} which, although still consistent with GR and a cosmological constant term, might require a new gravity theory or the existence of an exotic form of matter known as dark energy.

In \cite{S80} it was noticed that the local 1-loop corrections to the Einstein-Hilbert action are quadratic in curvature, and lead to a period of inflation in the very early universe. This motivates the introduction of additional functions of the curvature invariants, that allow for a dynamical late time acceleration \cite{C02, S07} without the need for additional scalar fields or cosmological constant $\Lambda$.
The observed value of the cosmological constant $\Lambda_\mathrm{obs}\simeq (10^{-4}\,\mathrm{eV})^4$ is so extraordinarily small that one needs a finely tuned bare cosmological constant $\Lambda_\mathrm{obs}=\Lambda_\mathrm{bare}+\Lambda_\mathrm{1-loop}+\Lambda_\mathrm{vev}$ in the Einstein Hilbert action in order to cancel the large quantum vacuum energy $\Lambda_\mathrm{1-loop} \geq E_\mathrm{ew}^4 \simeq (100\,\mathrm{GeV})^4$ and classical contributions related to Standard Model phase transitions $\Lambda_\mathrm{vev}\simeq\mathcal{O} (E_\mathrm{ew}^4,E_\mathrm{QCD}^4)$. This severe ``old'' cosmological constant problem \cite{W89} is not resolved in $f(R)$ theories.
 In this paper we assume that effectively $\Lambda_\mathrm{bare}=\Lambda_\mathrm{1-loop}=\Lambda_\mathrm{vev}=0$ due to some other physical process acting at the scale $\Lambda_\mathrm{obs}\simeq (10^{-4}\,\mathrm{eV})^4$ such that the scale $\Lambda_\mathrm{obs}$ enters the  $f(R)$ function naturally.
In addition, the unknown physics leading to the removal of the vacuum energy might be accompanied by an effective scalar degree of freedom.
There exist some ideas of screening \cite{DHK07, deRetal08, BDHKM11} and relaxing \cite{D85, CCPS11, EK11, BSS10} the vacuum energy. 

In this work we focus on a simple class of modified gravity models, so called $f(R)$ gravity, where an arbitrary function of the Ricci scalar is introduced into the standard Einstein Hilbert action. It is well known that certain $f(R)$ functional forms can give rise to an expansion history that exactly mimics a Universe governed by a cosmological constant $\Lambda$ and cold dark matter ($\Lambdaup$CDM). However even for such models, the modified gravity contribution will still affect the growth of structure, both in the linear and nonlinear regime via a fifth force mediated by a scalar degree of freedom, the ``scalaron'' \cite{S80} . 

To be consistent with local and astrophysical gravity tests, $f(R)$ functions must be used that essentially approach a constant value (usually $\Lambda_\mathrm{obs}$) in regions of high curvature. This ensures that in these high curvature regions the local scalaron mass becomes large enough to shut down the fifth force on much smaller scales. This effect is called the chameleon mechanism \cite{KW04a,BvBDKW}. At high curvature, such as in regions with deep potential wells in the late universe, e.g.\,galaxies and the solar system or in the early universe, at recombination, these $f(R)$ theories are pushed towards GR. However at small curvature, typically well after matter-radiation-equality and on large scales, detectable deviations from GR are possible and still allowed observationally. This is exactly the regime where linear and non-linear large scale structure formation takes place. While $f(R)$ theories should not be viewed as fundamental theories of gravity, they offer a self-consistent tool to scrutinize GR and look for new physics in the currently mapped large scale structure.

In \cite{OLH08,MMMS11} a constraint on $f(R)$ theories using the galaxy power spectrum was calculated , where it was found that the chameleon mechanism supresses deviations from GR on small scales where perturbations become nonlinear. However the formation of galaxy clusters involves both linear and nonlinear growth and is thus better suited to probe both the fifth force and the chameleon effects present in scalar-tensor theories \cite{LH11}.

The aim of this paper is to predict the number density of dark matter halos $n(M,z,f_{\rm R0})$ with a given mass $M$ and observed redshift $z$ as a function of the model parameter $f_{\rm R0}$ of the Hu-Sawicki-Starobinsky $f(R)$ model \cite{HS07,S07}.  Building on the work of \cite{CA1,achitouvetal}, we employ the spherical collapse model to obtain a critical density contrast $\delta_c$, from which we construct a realistic mass function. Similar semi-analytical formalisms were already applied to $f(R)$ theories in previous works \cite{SLOH09,BRS10,BJZ11,LE11,LL12b, LL12,LLKZ13}, see \cite{CCL13} for voids. In \cite{SLOH09} only two limiting cases were considered: either the collapsing overdensity was considered to be fully chameleon screened or fully unscreened. As these two cases correspond to spherical collapse in $\Lambdaup$CDM with a rescaled Newton constant in the unscreened case, an initial top-hat density profile retains its shape during collapse. This allowed the authors to study the spherical collapse of a top-hat profile by simply comparing the evolution of a closed patch of Friedmann-Robertson-Walker (FRW) spacetime. However comparison to $N$-body simulations showed that this model was too simple and missed the interesting regime between the two limiting cases \cite{SLOH09}. 

The above restrictions were dropped in \cite{BJZ11}, where the collapse of the top-hat was studied numerically by solving the modified gravity field equations. An important result of this work was the discovery that a top-hat profile develops shell crossing during its evolution. In order to alleviate this problem, a smooth transition region between the top-hat and the `background' FRW spacetime was introduced. However the resulting values of $\delta_c$ showed dependence on the shape of the transition region, and did not lie in the expected range obtained in  \cite{SLOH09}. 

In this paper we improve the spherical collapse calculation for $f(R)$ models by using as initial condition the average density profile around a density peak. This is completely determined by the input cosmology, thus removing any ambiguity in the choice of initial profile and making the spherically symmetric setup as physically accurate as possible. The profile is calculated using peaks theory \cite{BBKS86} and the linear matter transfer function \cite{LCL00}.

The paper is organized as follows. In Section \ref{section:equations} and \ref{subsec:Sphericalcollapse} we review and obtain the relevant equations for spherical collapse in $f(R)$ theories and explain the method of their numerical solution. Here special care is taken of the initial conditions and applicability of the quasistatic approximation, which is crucial due to the breakdown of Birkhoff's theorem.  We exhibit one of the main results of this work; a fitting function for the critical overdensity $\delta_c(M,z,f_{R0})$ which the linearly extrapolated matter density has to reach in order to form a halo of mass $M$ at redshift $z$ within the model parameter range $10^{-7}<f_{\rm R0}<10^{-4}$. In Section \ref{section:massfunction} we review the excursion set formalism and extending $\delta_c(M,z,f_{R0})$ into a drifting and diffusing barrier, modeling this way aspherical collapse, in order to obtain a realistic mass function $n(M,z,\delta_{c},f_{R0})$ in terms of the collapse redshift $z$, mass of the object $M$ and modified gravity parameter $f_{R0}$. The accuracy of this mass function function is checked against Monte Carlo random walks in Section \ref{section:massfunction} and we conclude in Section \ref{section:conclusion}.

\section{Review of the Hu-Sawicki-Starobinsky $\mathbold{f(R)}$ model}\label{section:equations}

\noindent The $f(R)$ action is given by
\begin{equation}
S= S_m+  \frac{1}{2\kappa^2}\int d^4 x \sqrt{-g}\, (R+f(R))\,,\label{fRAction}
\end{equation}
where $\kappa^2=8 \pi G$ and $S_m$ a minimally coupled matter action. Variation with respect to the metric gives the Einstein field equations
\begin{equation}
G_{\mu \nu}=  e^{-\varphi}\left(\kappa^2 T_{\mu \nu}-\frac{1}{2}g_{\mu \nu} V(\varphi) + (e^{\varphi})_{;\mu \nu}-g_{\mu \nu} \Box e^{\varphi} \right)\,, \label{Einsteineq}
\end{equation}
with energy momentum tensor $T_{\mu \nu}$ and Einstein tensor $G_{\mu \nu}=R_{\mu \nu}-R g_{\mu \nu}/2$. We have introduced the notation
\begin{equation}
e^{\varphi}\equiv 1+ f_{,R}\,, \qquad V(\varphi)\equiv R e^{\varphi} -R-f\,,\label{phidef}
\end{equation}
where writing $V(\varphi)$ requires a form of $f$ such that $e^{\varphi}= 1+ f_{,R}$ can be inverted to give $R(\varphi)$. The condition $1+ f_{,R}>0$ is required to ensure that the $f(R)$ model remains ghost free \cite{S07}.
Using \eqref{phidef}, the trace of the field equations 
\begin{equation}
3 \Box e^{\varphi} = -2 V + V_{,\varphi}+\kappa^2 T\label{phieom}
\end{equation}
\noindent allows us to interpret the fourth order differential equations for $g_{\mu \nu}$, Eqs.\,\eqref{Einsteineq} and \eqref{phidef}, as second order field equation \eqref{Einsteineq} plus a second order equation \eqref{phieom} for the scalar $\varphi$. One can view this procedure as a first step in the direction of a Hamiltonian formulation \cite{OS11, SH11, DSY09}.
It is also possible to make the replacement \eqref{phidef} already in the action \eqref{fRAction}
\begin{equation}
S= S_m+  \frac{1}{2\kappa^2}\int d^4 x \sqrt{-g}\, (e^{\varphi} R-V(\varphi))\,,\label{phiAction}
\end{equation}
where variation with respect to $g_{\mu \nu}$ and the scalar $\varphi$ leads to \eqref{Einsteineq} and \eqref{phidef}.
The action \eqref{phiAction} is kown as O'Hanlon theory \cite{OH72}. Since matter is minimally coupled, the energy momentum tensor of matter $T_{\mu \nu}$ is conserved
\begin{equation}
T^{\mu}{}_{\ \nu;\mu}=0\,,
\end{equation}
\noindent implying that the Euler and continuity equation for a perfect fluid take the same form as in GR. The same is true for the geodesic equation and thus for the general relativistic virial theorem \cite{J70}.\footnote{Eq.\,(8) of \cite{J70} applies unchanged to metric $f(R)$ theories. In order to estimate the virial radius within the context of spherical collapse one needs in addition to the virial theorem some form of energy conservation. This however does not exist in general $f(R)$ theories \cite{CCHO09}.} In the following we consider only cold dark matter in the single stream approximation, such that the energy momentum tensor takes the form of pressureless perfect fluid
\begin{equation}
T_{\mu \nu}=\rho u_\mu u_\nu\,,\quad u_{\mu} u^{\mu}=-1\,,\quad T=-\rho\,,
\end{equation}
with energy density $\rho$ and four velocity $u^\mu$. This is justified as we only consider times well after matter radiation equality and choose initial conditions where shell crossing does not occur. 

The functional form of $f(R)$ is strongly restricted due to theoretical and phenomenological constraints \cite{HS07, DeFT10}, meaning that the resulting theory is free of classical and quantum instabilities and consistent with all known gravity experiments and observations.
We will consider the class of $f(R)$ functions proposed by Hu and Sawicki \cite{HS07} and also Starobinsky \cite{S07}
\begin{align}
f(R)=&-2\Lambda+ \frac{\epsilon}{n} \frac{( 4\Lambda)^{n+1}}{R^n}\label{powerlaw}\\ 
\simeq&-  \frac{2\Lambda}{1+2 \epsilon/n\ (4\Lambda/R)^n} \label{brokenpowerlaw}\\
\simeq&- 2\Lambda +2\Lambda \left(1+ \left[ \frac{(n/2)^{1/n}}{\epsilon}  \frac{R}{4\Lambda}\right]^2 \right)^{-n/2} \,,\label{brokenpowerlawStaro} 
\end{align}
which can fulfill the above-mentioned constraints and are thus viable. In the cosmologically important region $R>4\Lambda$ the three models above exhibit essentially the same behaviour. The Hu-Sawicki model (\ref{brokenpowerlaw}) and the Starobinsky model (\ref{brokenpowerlawStaro}) interpolate between $f=0$ and $f=-2 \Lambda$, where steepness is controlled by $n$ and the position of the transition is at $\epsilon^{1/n}4\Lambda.$ Here $\Lambda$ is a constant energy scale whose value coincides with the measured value $\Lambda=\Lambda_\mathrm{obs}=3 H_0^2 \Omega_\Lambda$ and $\epsilon \ll 1$ is a small positive deformation parameter which is related to the more commonly used $f_{\rm R0}$, via
\begin{equation}
f_{\rm R0} \equiv |f_{,R}(R_0)|=  \epsilon \left( 1+ \tfrac{1}{4}\left(\Omega_\Lambda^{-1}-1\right) \right)^{-(n+1)}\,.
\end{equation}
Galactic rotation curves require $\epsilon \lesssim 10^{-6}$ and cepheids $\epsilon \lesssim 0.5\times 10^{-6}$, such that viable $f(R)$ theories and $\Lambdaup$CDM have virtually indistinguishable expansion histories \cite{HS07, JVS12}. Stability of the de Sitter vacuum together with solar system tests demands $n\gtrsim 1$ \cite{CT08}. Using the expansion history and baryon acoustic oscillations (BAO) gives complementary constraints on $n$ \cite{MMMS11}. Note also that quantum corrections to the scalaron mass place strong constraints on $\epsilon$ \cite{UHK12}.
Enforcing $\epsilon\ll 1$ effectively introduces two additional energy scales; $ \Lambda/\epsilon$ and $\Lambda \epsilon$. As we will see, $\Lambda \epsilon$ is the range over which the effective potential of $\varphi$ varies and  $\Lambda/\epsilon$ is the squared mass of the scalaron, corresponding to small fluctuations ($\ll \epsilon$) around the background field $\bar{\varphi}=-\epsilon$ during the cosmic late time acceleration.

\begin{figure}[t] 
\centering
\includegraphics[width=0.45 \textwidth]{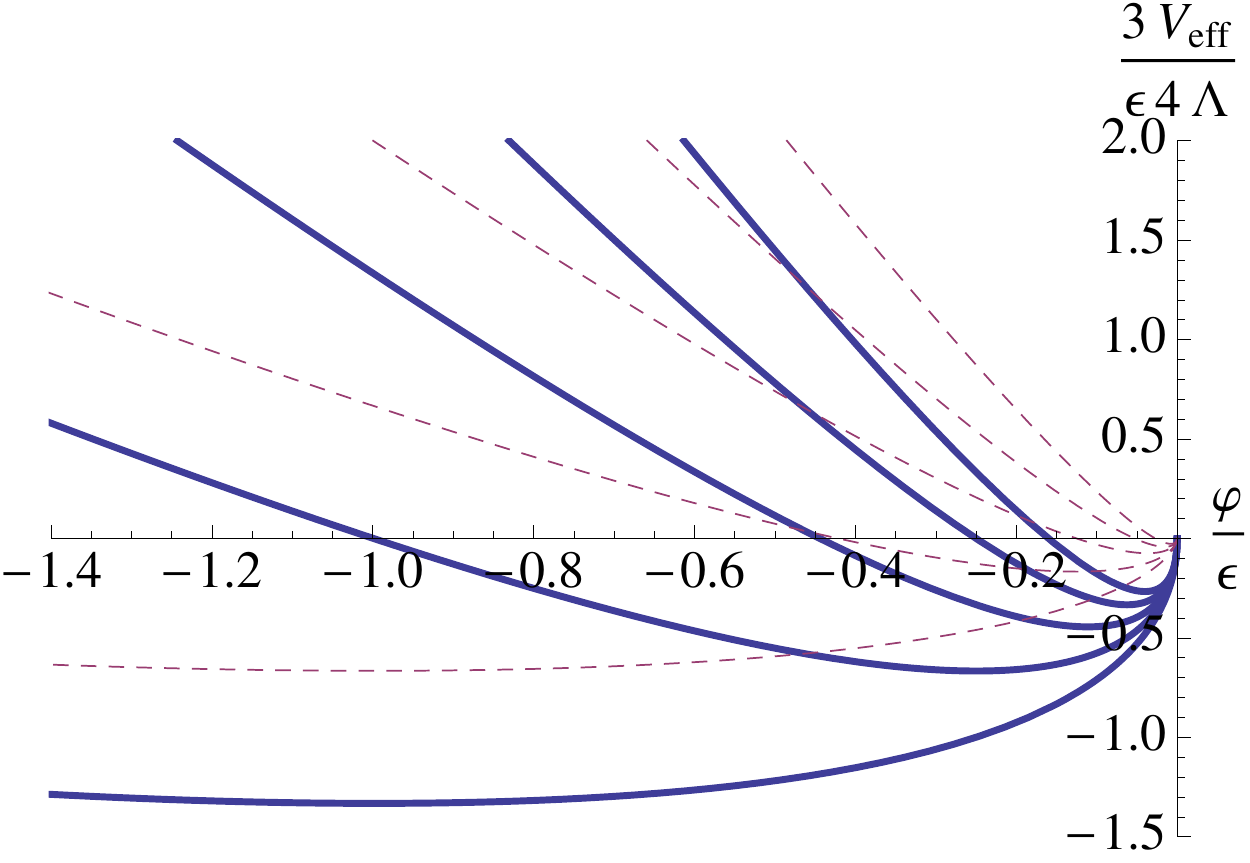}
\caption{$V_\mathrm{eff}$ for model \eqref{powerlaw} with $n=1$, {\it full} lines, and $n=2$, {\it dashed} lines. From bottom to top $\kappa^2 \rho/4\Lambda=10^{-3}, 1, 2, 3$ and $4$.
}
\label{fig:Veff}
\end{figure}

Now we derive the explicit form of $V(\varphi)$ for model \eqref{powerlaw}, which will be assumed in the rest of this paper. For this we note first that 
\begin{equation}
f_{,R}= - \epsilon \left(\frac{4\Lambda}{R}\right)^{n+1}\,, \label{frexpression}
\end{equation}
such that in and after matter domination 
\begin{equation}
|f_{,R}| \leq  \epsilon \ll 1\qquad \Rightarrow\qquad f_{,R}\simeq\varphi\,. \label{smallphi}
\end{equation}
From Eqs. \eqref{frexpression} and \eqref{smallphi} it follows
\begin{equation}
 R= 4\Lambda  \left(\frac{|\varphi|}{ \epsilon}\right)^{-\frac{1}{n+1}}\,, \qquad
f=-2\Lambda+ \frac{\epsilon}{n} 4\Lambda \left(\frac{|\varphi |}{ \epsilon}\right)^{\frac{n}{n+1}}\,.
\end{equation}
Finally $V=\varphi R-f$ becomes 
\begin{equation}
V=2\Lambda\left(1- 2 \epsilon \frac{n+1}{n} \left(\frac{|\varphi |}{ \epsilon}\right)^{\frac{n}{n+1}}\right)\,.
\end{equation}
We can write the scalar equation \eqref{phieom} as 
\begin{equation}
V^{\mathrm{eff}}_{,\varphi}\equiv-e^{-\varphi}\frac{1}{3}\left(2V+\kappa^2 \rho- V_{,\varphi} \right)=\square \varphi\,,
\end{equation}
where $V^{\mathrm{eff}}$ is the effective potential the scalar $\varphi$ tries to minimize. Using $|\varphi|\ll1$ it is given by
 \begin{equation}
V^{\mathrm{eff}}= \epsilon\frac{4\Lambda}{3} \left(\left(1 +\frac{\kappa^2 \rho}{ 4\Lambda}\right)\frac{|\varphi|}{\epsilon}  - \frac{n+1}{n} \left(\frac{|\varphi |}{ \epsilon}\right)^{\frac{n}{n+1}}\right)\,.
 \end{equation}
The minimum of $V^{\mathrm{eff}}$ is at the field value $\varphi_\mathrm{min}$,
\begin{equation}
\varphi_\mathrm{min}=-\epsilon \left(1 +\frac{\kappa^2 \rho}{ 4\Lambda}\right)^{-(n+1)}\,,\label{phimin}
\end{equation}
 \noindent which approaches zero $\varphi_\mathrm{min}\rightarrow 0$ for $\kappa^2\rho\gg\Lambda$, see Fig\,\ref{fig:Veff}.
If $\varphi$ occupies the minimum in this limit then GR is restored: the effective Newtonian constant $e^{-\varphi} \kappa^2 \rightarrow \kappa^2$ returns to its GR value and the potential becomes $V\rightarrow 2 \Lambda$. For small fluctuations around $\varphi_\mathrm{min}$, the scalaron has a mass 
\begin{equation}
m^2 \equiv V^{\mathrm{eff}}_{,\varphi \varphi}= \frac{4}{3(n+1)} \frac{\Lambda}{\epsilon}\left(\frac{|\varphi |}{ \epsilon}\right)^{-\frac{n+2}{n+1}}\,,\label{mass}
 \end{equation}
which diverges at the General Relativistic limit $m(\varphi=0) \rightarrow \infty$.\footnote{Adding a $R^2/6M^2$ to \eqref{powerlaw} removes this infinity amongst other pathologies \cite{ABS10}, but leaves the late time cosmology unaffected provided the energy scale $M$ is large enough.}  In this limit the fifth force is turned off since the effective interaction range of a Yukawa type interaction is given by the Compton wavelength $1/m$. 

Whether $\varphi$ actually sits at the potential minimum in high density regions depends on the magnitude of spatial gradients \cite{KW04b}. For galaxy clusters, if the forming cluster is too small or $\epsilon$ to large, $\varphi$-gradients cost too much energy and $\varphi$ will stay near its background value $\bar{\varphi}=\bar{\varphi}_{\rm min}$ even within the cluster, preventing the chameleon mechanism from operating. Large clusters and small $\epsilon$ on the other hand leave enough space for the scalar to change from $\bar{\varphi}$ to $\varphi^{\rm cluster}_\mathrm{min}$ and hence the chameleon mechanism becomes active above a certain cluster mass scale. A necessary condition for the chameleon mechanism to activate is that the Compton wavelength $1/m_\mathrm{min}$ must be much smaller than the size of the overdensity \cite{HS07}. In this case there is only a ``thin shell'' at the edge of the object that can mediate a large distance fifth force, from which the interior is unaffected and thus behaves like a typical overdensity in GR.

For overdensities of different magnitude and shape, and for different values of the $f(R)$ model parameters ($n,\epsilon$) we encounter a time dependent mixture of all the above cases.

\section{Spherical collapse}\label{subsec:Sphericalcollapse}
The spherical collapse model \cite{P67, GG72} is a deterministic criterion which allows one to map 
an initially small, spherically symmetric overdensity to the formation of a virialized dark matter halo. More concretely this spherical collapse of an initial density profile $\delta(z_i,r)$ allows one to estimate the formation time $z_c(\delta_i)$ of a halo as function of the initial density amplitude $\delta_{i}\equiv\delta(z_i,r=0)$, which can be inverted to give the threshold 
\begin{equation}
\delta_i(z_c) \equiv \delta(z_i,r=0|z_c) \label{defdelc}
\end{equation}
for the initial density profile to collapse at redshift $z_c$.
Spherical collapse in an Einstein-de Sitter or $\Lambdaup$CDM universe can be modeled analytically by assuming that the density is homogeneous within the perturbation. Due to Birkhoff's theorem, the inner part is not influenced by the transition region and  simply behaves as a closed FRW universe. The redshift $z_c$ of collapse of this patch measured in the flat exterior FRW then approximately equals the formation time of a bound virialized object \cite{Jetal01}. 

In the case of $f(R)$ theories we actually need to solve the full field equations since Birkoff's theorem does not apply. There are further complications which hamper the calculation in $f(R)$ models. It was noticed in \cite{LH11} that halos are actually composed of subhalos, and this will increase the chameleon effect as screened subhalos attract each other less strongly than particles in a homogeneous dust cloud. Another environmental effect arises due to the fact that the forming cluster is itself a subcluster of a larger sized over/underdensity, enhancing/diminishing the chameleon effect  \cite{LE11,LLKZ13} by diminishing/enhancing field gradients. In this work we do not consider these two effects, although part of the environmental dependence of the collapse threshold is taken into account by using the average density profile around a peak, which only depends on linear power spectrum $P(z_i,k)$. We comment further on the environment in \ref{sec:den} and \ref{section:conclusion}.

As will be reviewed in Section \ref{section:massfunction} the halo mass function depends on the spherical collapse derived quantity $\delta_i(z_c)$ via
\begin{equation} \label{nudef}
\nu_c\equiv \frac{\delta_i(z_c)}{\sigma(z_i,R)}\,,
\end{equation}
which is $z_c$-independent for an Einstein de Sitter universe and slightly $z_c$-dependent in an $\Lambdaup$CDM universe. The standard deviation $\sigma(z_i,R)$ is given by the linear matter power spectrum $P(z_i,k)$ and filter function $W$
\begin{equation}
\sigma(z_i,R)^2= \int \frac{d^3 k}{(2\pi)^3} W(kR)^2 P(z_i,k)\,.
\end{equation}
For convenience one usually considers 
\begin{equation}
\delta_c(z_c) \equiv  D(z_c,z_i) \delta_i(z_c)\,, \label{deltaclinex}
\end{equation}
which defines the collapse threshold at redshift $z_c$. This quantity is the linearly extrapolated density field, where the linear growth function $D(z_c,z_i)= D(z_c)/D(z_i)$ was used to evolve from $z_i$ to $z_c$. The introduced time evolution has no physical meaning but is convenient as $\delta_c(z_c)=1.686$ is constant in an Einstein-de Sitter universe, and only weakly dependent on $z_c$ in $\Lambdaup$CDM. The approximate $z_c$- and $R$-independence of $\delta_c$ leads to the universality of halo mass function if written as a function of $\sigma(z_c,R)=D(z_c,z_i)\sigma(z_i,R)$. Despite the artificial time evolution introduced in the definition of \eqref{deltaclinex}, it is the collapse criterion \eqref{defdelc} defined at the initial time that one should have in mind both for GR and $f(R)$ gravity: the halo mass function is determined within the initial conditions and the information about formation time only enters via \eqref{defdelc}. Due to the practically identical expansion histories in $\Lambdaup$CDM and the assumed $f(R)$ model \eqref{powerlaw}, the initial conditions can be assumed to be equivalent for both models. In particular $\sigma(z_i,R)$ is identical in both models if $z_i$ is chosen such that all relevant scales are still linear. Therefore in Eq.\,\eqref{nudef} only $\delta_i(z_c)$ should be adjusted when we consider $f(R)$ models.

Since the collapse criterion \eqref{defdelc} is the quantity we wish to calculate, we can trivially rewrite the definition \eqref{nudef} of $\nu_c(z_c,R)$ using the $\Lambdaup$CDM growth function $D$
\begin{equation}
\nu_c\equiv \frac{\delta_i(z_c,R)}{\sigma(z_i,R)}=
\frac{\delta_i(z_c,R)D}{\sigma(z_i,R)D}=\frac{\delta_c(z_c,R)}{\sigma(z_c,R)}\,. \label{nucinfofr}
\end{equation}
\noindent This expression holds for both $f(R)$ and $\Lambdaup$CDM, however as we will see in Section \ref{section:massfunction}, $f(R)$ theories predict a $\delta_c$ that is a function of both $z_c$ and $R$. This is already the case for ellipsoidal collapse in GR \cite{ST}.

As a final point, let us emphasize that using the scale dependent $f(R)$ linear growth factor instead of the GR equivalent $D(z)$ in Eq.\,\eqref{nucinfofr} would be both incorrect and inconvenient. Incorrect since linear growth in $f(R)$ is scale dependent and therefore not multiplicative, and inconvenient as $\delta_c(z_c,R)$ would no longer encode all of the $f(R)$-dependent deviations; instead one would need to provide $\nu_c(z_c,R)$. $\delta_c(z_c,R)$ as defined in (\ref{defdelc}) and (\ref{deltaclinex}) is a convenient way to provide $\nu_c(z_c,R)$. Another convenient way would be to fix $\delta_c(z_c,R)\equiv \delta^{\Lambdaup}_c(z_c)$ and to fold all the $f(R)$-dependence into a modified $\sigma^{f(R)}(R)$ \cite{LH11}.

\subsection{Quasistatic equations}
To obtain $\delta_c$, we must calculate the collapse of a spherically symmetric pressureless matter distribution in an asymptotic FRW spacetime, such that the 3+1 dimensional problem simplifies to a 1+1 dimensional one. In GR the calculation is much simpler. Due to Birkhoff's theorem an initially homogeneous (``top-hat'') overdensity retains its shape during collapse. This allows us to treat the size of the homogeneous overdense region as the scale factor of a closed FRW universe \cite{We08}.

In $f(R)$ theories the additional scalar degree of freedom $\varphi$ allows for monopole radiation \cite{S66}, thus Birkoff's theorem no longer applies. Another, more severe, problem is that in the linear regime of collapse the gravitational force is scale dependent due to mass of $\bar{\varphi}$ fluctuations. Finally, since the energy density becomes sufficiently large during collapse for the chameleon mechanism to take effect, the gravitational force will depend on the local density. As a result of these effects, an initial top-hat overdensity will not retain its shape during collapse and we cannot use a closed FRW to describe its collapse. Rather, we must solve the spherically symmetric $f(R)$ field equations. 

A spacetime which a admits a spherically symmetric spatial slicing has a metric which can written in the form \cite{MTW73}
\begin{equation}
d s^2= - e^{2 \Phi} d t^2 + a^2 e^{-2 \Psi}(dr^2 + r^2 d\Omega^2)\,,\label{spherMet}
\end{equation}
where both $\Phi$ and $\Psi$ are functions of $r$ and $t$. For convenience we factor out $a(t)$, which will be the scale factor of the asymptotic flat FRW spacetime, where we choose $a(t_0)=1$ without loss of generality. Note that this metric is fully nonlinear. We present the nonlinear field equations in Appendix \ref{appendix:nonlineareq} and derive the conditions under which one can assume that $\Phi$, $\Psi$ and $\varphi$ remain small even when the density becomes non-linear $\delta\equiv \rho/\bar{\rho}>1$.
Under these conditions, the set of relevant field and fluid equations reduces to the

\begin{subequations}\label{fofRequationslinfinal2}
\noindent Poisson equation
\begin{equation}
a^{-2}\lapl \Phi= \frac{2}{3}\kappa^2 \bar{\rho}\delta-\frac{1}{6} \delta\!V_{,\varphi}\,, \label{Poissoneq}
\end{equation}
\noindent the nonlinear scalar field equation
\begin{equation}
a^{-2}\lapl \varphi=\frac{1}{3}(\delta\!V_{,\varphi}-\kappa^2 \bar{\rho}\delta)\,,\label{phieq2}
\end{equation}
\noindent the energy conservation
\begin{equation}
\dot{\delta} + \frac{1}{ar^2}\partial_r\left( r^2(1+\delta) v\right)=0
\end{equation}
\noindent and the Euler equation
\begin{equation}
\dot{v}+v H+\frac{v}{a}v'= -\frac{1}{a}\Phi'\,,\label{eulereq}
\end{equation}
\end{subequations}
with $v\equiv v^r\equiv a u^r/u^0$ as the radial velocity and $\delta\!V_{,\varphi}\equiv V_{,\varphi}-\bar{V}_{,\varphi}$ the perturbation in the Ricci curvature. Note that we cannot assume $\delta\!V_{,\varphi}= \bar{V}_{,\varphi \varphi} \delta \varphi$ in  \eqref{phieq2} since we would then miss the effect of the chameleon mechanism. It is important to treat  \eqref{phieq2} as a nonlinear partial differential equation, even though $\delta \varphi \equiv \varphi -\bar{\varphi} \leq \epsilon \ll 1$. This non-linearity makes solving the system \eqref{fofRequationslinfinal2} a nontrivial task. We explain details of the numerical methods in Section \ref{section:numerics}.

\subsection{Initial conditions} 
In addition to the relevant dynamical equations, we must also specify the initial conditions of the problem. Well after matter-radiation equality and well before the late time accelerated expansion, the Universe is in a state that is well described by a linearly perturbed Einstein-de Sitter spacetime on all scales relevant for large scale structure formation. At such early times modified gravity effects due to the scalar field $\varphi$ are completely negligible due to the temporal chameleon effect as seen in Fig.\,\ref{fig:Veff}.

We choose $z_i=500$ as our initial time for the spherical collapse. At this redshift, radiation is already sub-dominant relative to matter by a factor of order $\sim {\cal O} (0.1)$.   Also the subhorizon assumption implicit in eq. \eqref{grPoisson} at $z_{i}$ is acceptable; the largest masses considered in this work are $\sim 0.1$ of the horizon size.\footnote{The largest clusters have masses $\sim 10^{15}\,M_\odot$ and they entered the horizon $ a H/k=1$ at around $a=3\cdot 10^{-5}$. In order to estimate the sizes at $z_i=500$ and today we assume a matter dominated universe, thus $H a\sim a^{-1/2}$ and we find $a H/k|_{z=0} \sim {\cal O} \left(10^{-3}\right)$ and $a H/k |_{z=500} \sim {\cal O} (0.1)$, which is well inside the horizon.} Note that at this initial redshift, the radiation component is subdominant to matter but non-negligible. To evade any potential problems with normalisation of the power spectrum, we use CAMB \cite{LCL00} to obtain $\sigma(z\!\!=\!\!0,R)$ with the choice of cosmological parameters \begin{align*}
\sigma_8&=0.8\,,\\
\Omega_m&=0.27\,,\\
h&= 0.7\,,\\
n_s&= 0.96
\end{align*}
and then evolve the general relativistic growth equation to obtain $\sigma(z_{c},R) = D(z_{c}) \sigma(z\!\!=\!\!0,R)$ to the collapse redshift, neglecting radiation. Throughout this paper, we stress that all quantities calculated using linear theory are obtained from the standard general relativistic equations.  

We discuss our choice of the initial density profile $\delta_{i}(r)$  in the following section. Given $\delta_{i}(r)$, we can use the constraint and Poisson equations
\begin{equation}
2 \Phi' H= -\kappa^2\bar{\rho}(1+\delta) a v \label{fofRmomconstrlinfinal2app}
\end{equation} 
\begin{equation}
2 a^{-2}\lapl \Phi= \kappa^2 \bar{\rho}\delta \label{grPoisson}
\end{equation}
 \noindent to obtain $\Phi_i(r)$ and $v_i(r)$ at $z=z_{ i}$;
\begin{equation}
v_i(r)=-\frac{a_iH_i}{r^2}\int_0^r \delta_i(r') {r'}^2 dr' \label{initialvel}\,.
\end{equation}

\noindent We use the the following natural boundary conditions at all times; $\Phi' = \Psi' = \varphi' = 0$ at $r=0$, and $\Phi = \Psi = 0$ at the outer boundary $r \to \infty$.

\subsection{Density profile}
\label{sec:den}

\begin{figure}[t] 
\centering
\includegraphics[width=0.49 \textwidth]{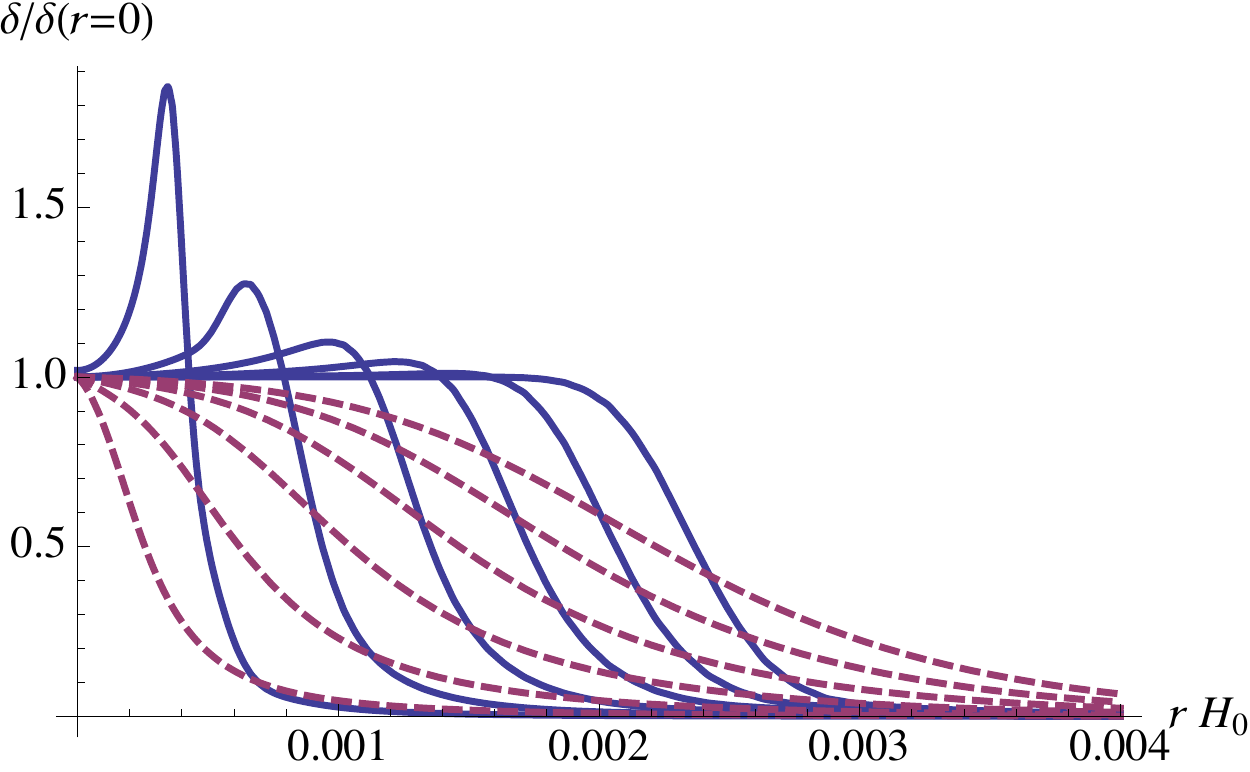}
\caption{Comparison of density profiles \eqref{tanhprofile} with  $s=0.05$,  {\it full} lines, and $s=0.4$, {\it dashed} lines for $n=1$ and $\epsilon=10^{-5}$. The plot shows the normalized density profile at different instances during collapse, from left to right $a=0.24,0.34,0.44,0.54,0.64,0.73$.
}
\label{fig:shells}
\end{figure}

It was observed in \cite{BJZ11} that the shape of an initial tophat density profile will evolve in $f(R)$ theories, in contrast to the shape preserving evolution obtained in GR. Specifically, they found that a tophat profile develops a large spike near the boundary between the overdensity and background FRW spacetime. We confirmed this behaviour when using the initial density profile
\begin{equation}
\delta_i(r)=\frac{\delta_{i,0}}{2}\left(1-\tanh\left(\frac{r/r_b-1}{s}\right)\right)\,,\label{tanhprofile}
\end{equation}
where $r_b$ is the size of the tophat-like function and $0<s<1$ determines the steepness of the transition, with $s\rightarrow0$ leading to $\delta_i(r)=\delta_{i,0} \theta(r-r_b)$.
Decreasing $s$ has the effect of forming a steeper spike at an earlier redshift. Fig.\,\ref{fig:shells} shows the normalized density profiles for two different steepness parameters but same $r_b$ at different instances during collapse.

\begin{figure*}[t] 
\centering
\includegraphics[width=0.32 \textwidth]{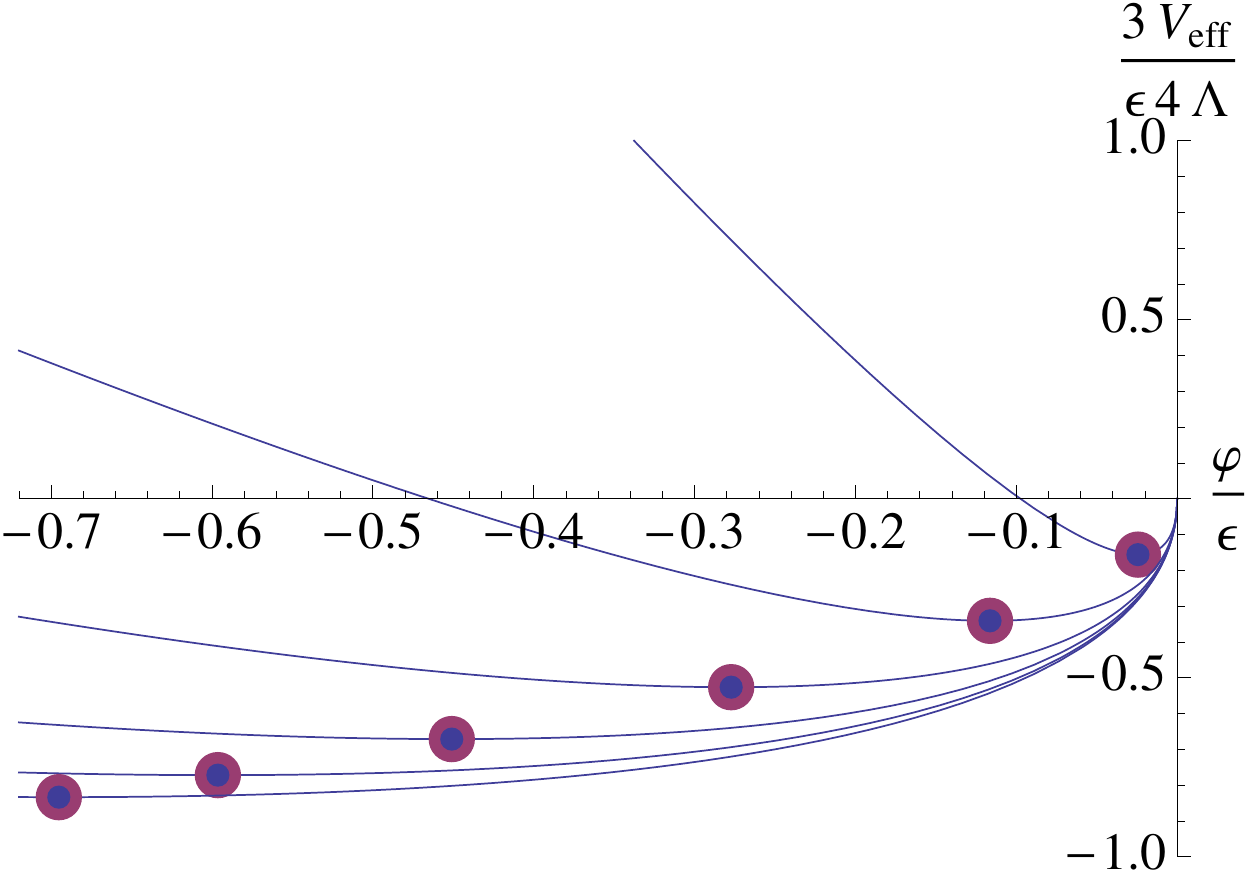}
\includegraphics[width=0.32 \textwidth]{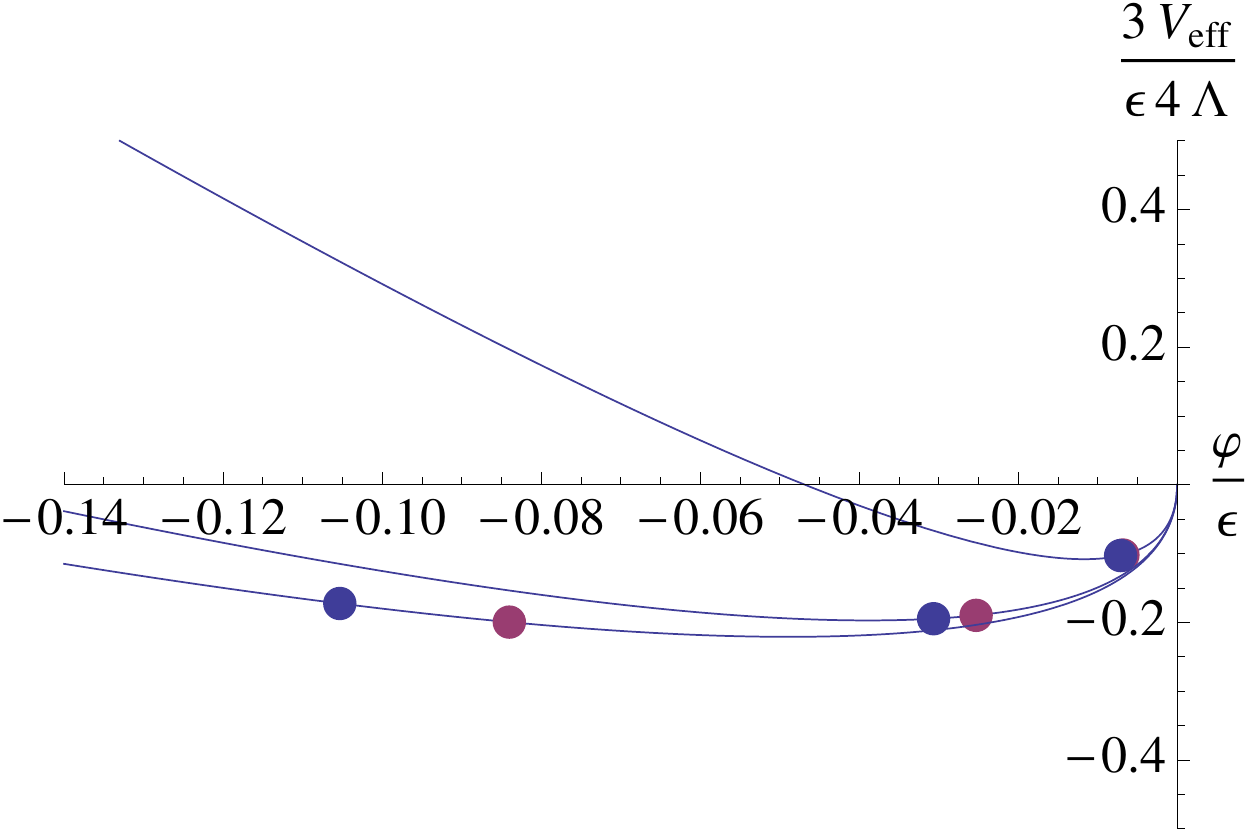}
\includegraphics[width=0.32 \textwidth]{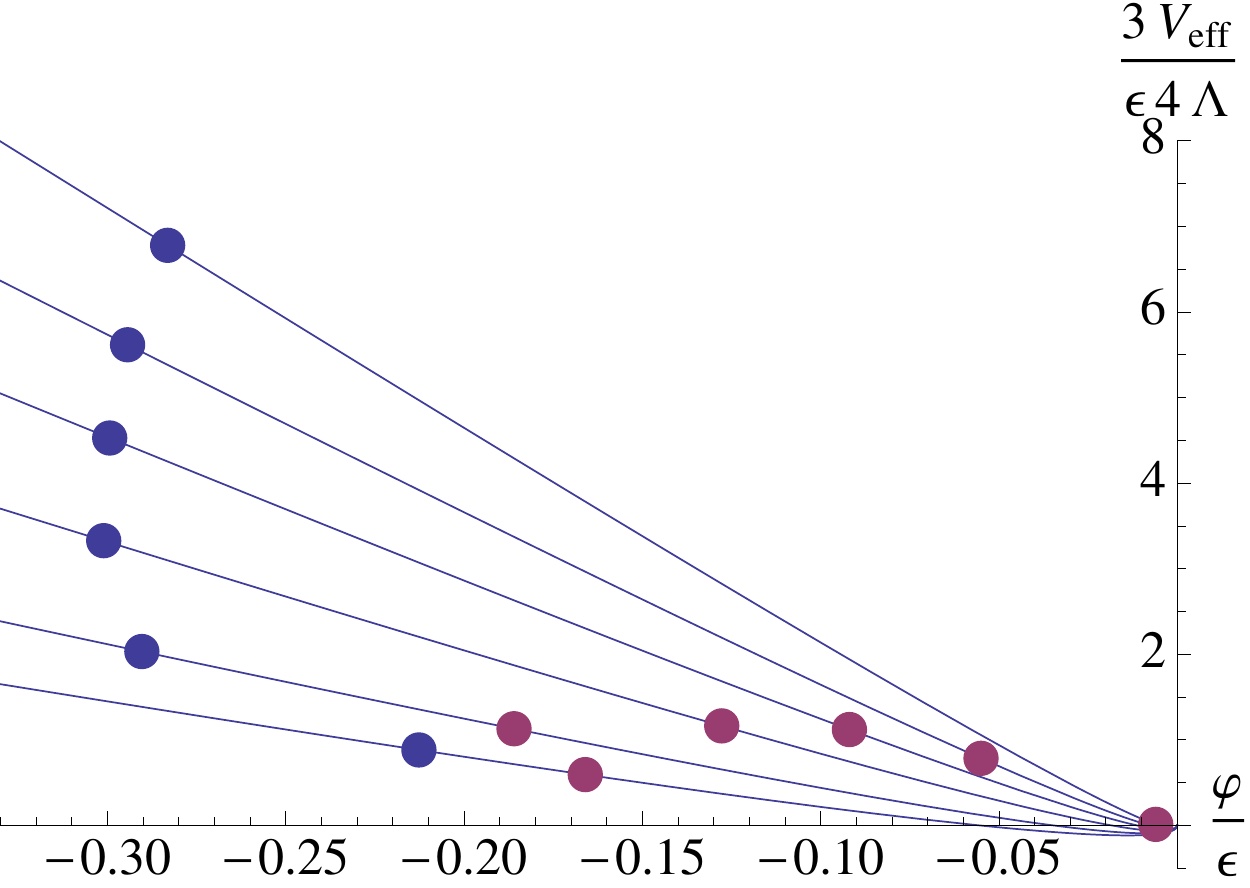}
\caption{Comparison of $\varphi(r)$  for density profiles \eqref{tanhprofile} with $s=0.05$,  {\it blue} dots, and $s=0.4$, {\it red} dots using $n=1$ and $\epsilon=10^{-5}$. 
In the left plot the effective potential $V_\mathrm{eff}$  ({\it full} line) is evaluated for $\varphi(a,r H_0=1)$ at different instances during collapse, from top to bottom $a=0.24,0.34,0.44,0.54,0.64,0.73$.
The middle and right plot show $V_\mathrm{eff}$ evaluated at the center $\varphi(a,r=0)$ at $a=0.24,0.34,0.44$ from top to bottom and $a= 0.54,0.64,0.69,0.72, 0.73$ from bottom to top.
}
\label{fig:chameleonshell}
\end{figure*}

\begin{figure}[t] 
\centering
\includegraphics[width=0.45 \textwidth]{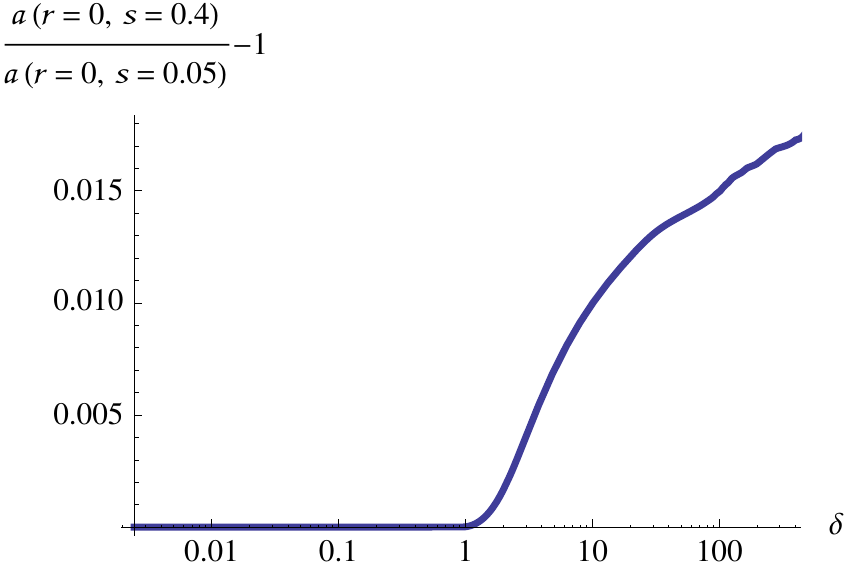}
\caption{Ratio of of the scale factor written as function of the central overdensities for $s=0.05$ and $s=0.4$ using the density profile \eqref{tanhprofile}  ($n=1$ and $\epsilon=10^{-5}$)}
\label{fig:deltaratio}
\end{figure}

The formation of a spike, which signals shell crossing, prohibits the use of tophat like functions for numerical studies. More importantly, it is clear that the shape of the density profile can dictate whether the chameleon mechanism becomes active; this is indicated in Fig\,.\ref{fig:chameleonshell}. Depicted here are the potential $V_{\rm eff}$ and field $\varphi$ at different times and positions. We see that the minimum of the effective potential $\varphi_{\rm min}$ only determines the position of $\varphi$ far outside the overdensity $r H_0=1$, where gradients are always small. In the center $r=0$ it is prevented from settling into the minimum because the necessary field gradients are energetically too costly. However for the steep profile (blue dots), $\varphi(r=0)$ finally turns around and moves to the potential minimum. Hence the collapse time of an overdensity with fixed $M$ and $\delta_i$ will depend on the shape of the density profile. Fig.\,\ref{fig:deltaratio} shows that the growth rates of the two profiles start to deviate once $\delta$ becomes nonlinear. While the scalaron is nearly screened for $s=0.05$, it remains unscreened for $s=0.4$, enhancing the growth. It is thus clear that the collapse time $z_c$ depends on the shape of the initial profile.

\begin{figure}[t] 
\centering
\includegraphics[width=0.45 \textwidth]{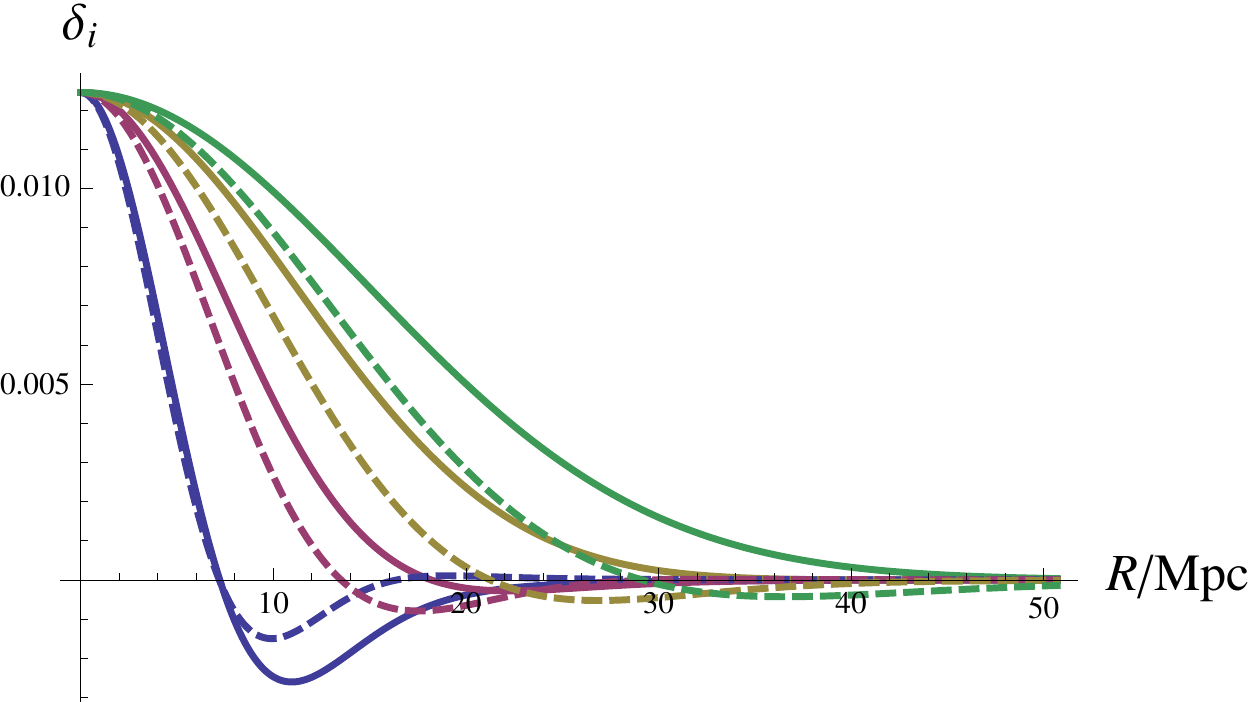}
\caption{Comparison between primordial ({\it dashed} lines) and transferred  ({\it full} lines)  density profile \eqref{peaksshape} for $\nu=1.5,2,2.5,3$ from left to right corresponding to smoothing lengths $R=11.1,8.6,6.0,3.9\,{\rm Mpc}/h$ is the smoothing length.}
\label{fig:densprofile}
\end{figure}

Due to the above subtleties, in our numerical calculations we use a physically motivated mean density profile around a peak of height $\nu\equiv\delta_{i,0}/\sigma_i(R)$
\begin{equation}
\delta_i(r,R)=\langle \delta(z_i,\mathbf{x},R)| \mathrm{peak},\nu\rangle \label{meanshape}
\end{equation}
which is completely determined by the gaussian statictics of the smoothed linear density field $\delta(z_i,\mathbf{x},R)$ \cite{BBKS86}. In Appendix \ref{section:peaks} we derive and display the explicit shape function, and in Fig.\,\ref{fig:densprofile} we exhibit the function for various initial $\nu$ values but fixed $\delta_{i,0}$. 
\noindent 
The mass contained in a spherical tophat 
\begin{equation}
M=\frac{4 \pi}{3} \bar\rho_0 R^3\,,
\end{equation}
\noindent is used to define the mass $M$ of the final halo, where $\bar\rho_0=\bar \rho(z\!=\!0)$ is the dark matter density at the present time. 

The physical reason for the above mentioned shape dependence of spherical collapse is the same as the environment dependence taken into account in \cite{LE11,LL12, LL12b,LLKZ13}.  In both cases the effectiveness of the chameleon mechanism is influenced by size of gradients in $\varphi$, which in turn depends on the density profile and its environment. The mean density profile \eqref{meanshape} is an approximate way to take into account both effects: the actual mean shape close to the ``size'' $r\simeq R$ of the collapsing protohalo and its mean environment for  $r>R$.

\subsection{Numerical method}
\label{section:numerics}

To numerically evolve the system of equations we start from the initial time slice $z_i=500$ and evolve the energy conservation and Euler equations over a single timestep, where we use $e$-foldings $N \equiv \ln[a]$ as our time variable with a staggered leapfrog method to decompose the temporal and spatial derivatives. For sufficiently small $\Delta N$ and coarse grained radial coordinate (we define $\bar{R} = \ln [r/r_{\rm scale}]$ as the radial coordinate, with $r_{\rm scale} = 1\,\mathrm{Mpc}$) the simple finite difference scheme remains stable. Between $z_{i}$ and $z=10$, we set the timesteps to be relatively large; $\Delta N = 2 \times 10^{-3}$, but for $z < 10$ we refine the timesteps to $\Delta N = 2 \times 10^{-4}$ to ensure that we can accurately model the effect of the chameleon mechanism. 

Once we have evolved the fluid equations to the next timestep, we solve the $\varphi$ equation \eqref{phieq2} using a very similar relaxation algorithm as outlined in \cite{BJZ11}; decomposing the non-linear equation into discretized form and Taylor expanding around the previous timestep. By solving the resulting large, yet sparse matrix equation, we update the solution and repeat until convergence is achieved. Once the field $\varphi$ is calculated on the new timestep, its contribution as a source in the Poisson equation equation is evaluated, and the metric potential $\Psi$ is obtained by a simple numerical integration. The process is then repeated until collapse is reached. We cannot evolve the system of equations formally to collapse. Here we simply evolve our system to an arbitrary high value in the non-linear over density; $\Delta_{\rm cut-off} = 10^{4}$. Once the collapse redshift $z_c$ has been determined, we use Eqs.\,\eqref{defdelc} and \eqref{deltaclinex} to obtain $\delta_{c}$.

To test that the solution obtained with our code is accurate, we check (completely independently of the code) that the data output $\delta, \varphi, \Psi, v$ solves the redundant momentum constraint and Newtonian gauge equations as given in Eqs.\,($\ref{eq:t1},\ref{fofRspattracelepert}$). We also test our code by attempting to reproduce the standard General Relativistic values of $\delta_{c}(z)$ (by using the algebraic relation $R = -8\pi G T$ as opposed to solving the non-linear $f(R)$ equation). We find an error of less than $0.3\%$ in $\delta_c^{\rm GR}$ for collapsing objects in the redshift range $z_{ c} = (0,2.5)$. Finally, we test that the code is unaffected by modifying the number of points used in the Poisson equation integration, changing the asymptotic boundary and decreasing the timesteps by a factor of ten. All tests produced a deviation of less than $0.5\%$ in the resulting $\delta_{c}$. 

We perform over $1000$ runs of the code, varying over initial density $\delta_{i}$, field value $f_{\rm R0}$ and scale of the perturbation $R$, choosing $\delta_{i}$ such that the overdensity collapses between $z=(0,2.5)$ and $R$ such that the mass of the object lies in the range relevant to clusters, $M = (10^{13},10^{15})M_\odot$.

\subsection{Spherical collapse threshold for $\mathbold{f(R)}$ gravity} 
\label{section:deltac}
The main results of this work are the $f(R)$ collapse threshold $\delta_{c}$ and a realistic halo mass function $n(M)$ as a function of $f_{\rm R0}$, $z$ and $M$, which we present in the following sections. In Section \ref{section:deltac} we will obtain $\delta_c$ as a fit function to our numerical results, and in Section \ref{section:massfunction} we use this $\delta_{c}$ adding a drifting and diffusing barrier in the excursion set theory to obtain a realistic mass function $n(M)$. 

The threshold for collapse will be a non-linear function of the $f(R)$ model parameters $n$ and $f_{\rm R0}$, and also the initial density $\delta_{ i}$ (or correspondingly the collapse redshift $z_{c}$), and the mass of the overdensity $M$ (or equivalently the size of the overdensity, fixed by $R$ as discussed in section \ref{sec:den}). We wish to construct a fitting function for $\delta_{c}$ using as input data the $N_{\rm code}=1000$ values of $\delta_{c}$ obtained from our numerical simulations. In what follows we fix $n=1$ for simplicity. 

\begin{figure*}[p!] 
\centering
\includegraphics[width=0.99 \textwidth]{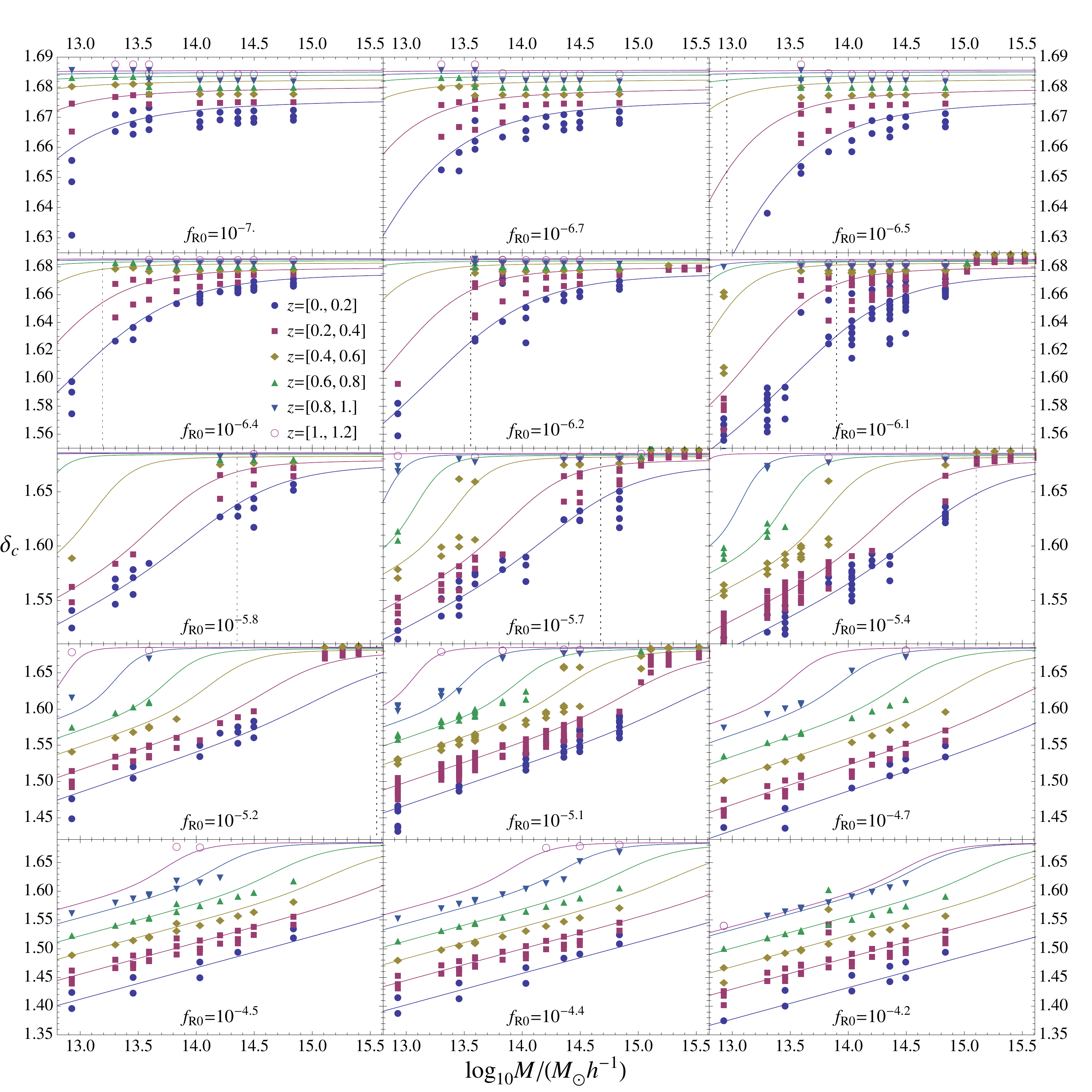}
\caption{$\delta_{c}$ as a function of $\log_{10}[M/(M_\odot h^{-1})]$. Each panel corresponds to spherical collapse runs with fixed $f_{\rm R0}$ and each set of data points of the same color/shape corresponds to runs which collapse within the same redshift bin, which can be inferred from the legend. The full lines show the fitting function Eq.\,\eqref{deltacfit} evaluated at the mean redshift within each of the redshift bins.  \vspace{5cm}}
\label{fig:deltacdetaandfit}
\end{figure*}

We exhibit the behaviour of $\delta_{c}$ as a function of $R$, $z_{c}$ and $f_{\rm R0}$ in Fig.\,\ref{fig:deltacdetaandfit}. These figures are instructive as a qualitative check of the influence of modified gravity on collapse. Each panel corresponds to runs with fixed $f_{\rm R0}$ and each set of data points of the same color/shape correspond to runs with the same average collapse redshift. We observe a clear linear dependence between $\delta_{c}$ and $\log_{10}[M/(M_\odot h^{-1})]$ for small $M$ and a redshift and $f_{\rm R0}$-dependent break in this behaviour. This break is determined by $m_b=0$ from eq.\,\eqref{deltacfit}. The dashed vertical line shows the mass as defined in \eqref{Mbreak} for which $m_b=0$ at $z=0$. We also observe an approach to the GR value $\delta^\Lambdaup_{c}$ for increasing $z_{c}$.  Similarly, for $f_{R0}$ field values close to the General Relativistic limit $f_{R0}=0$, we find the correct limit $\delta_{c} \to 1.686$. The full lines show the fit function \eqref{deltacfit}. 
We clearly observe a non-trivial $R$ and $z_{c}$ dependence and a return to GR for large objects and those with a high collapse redshift. An approximately linear relationship between $\delta_{c}$ and $\log[M/M_\odot]$ is observed for large values of $f_{\rm R0} \gtrsim 10^{-5}$, in agreement with the results of \cite{LE11, LLKZ13}. 

\begin{figure}[t] 
\centering
\includegraphics[width=0.45 \textwidth]{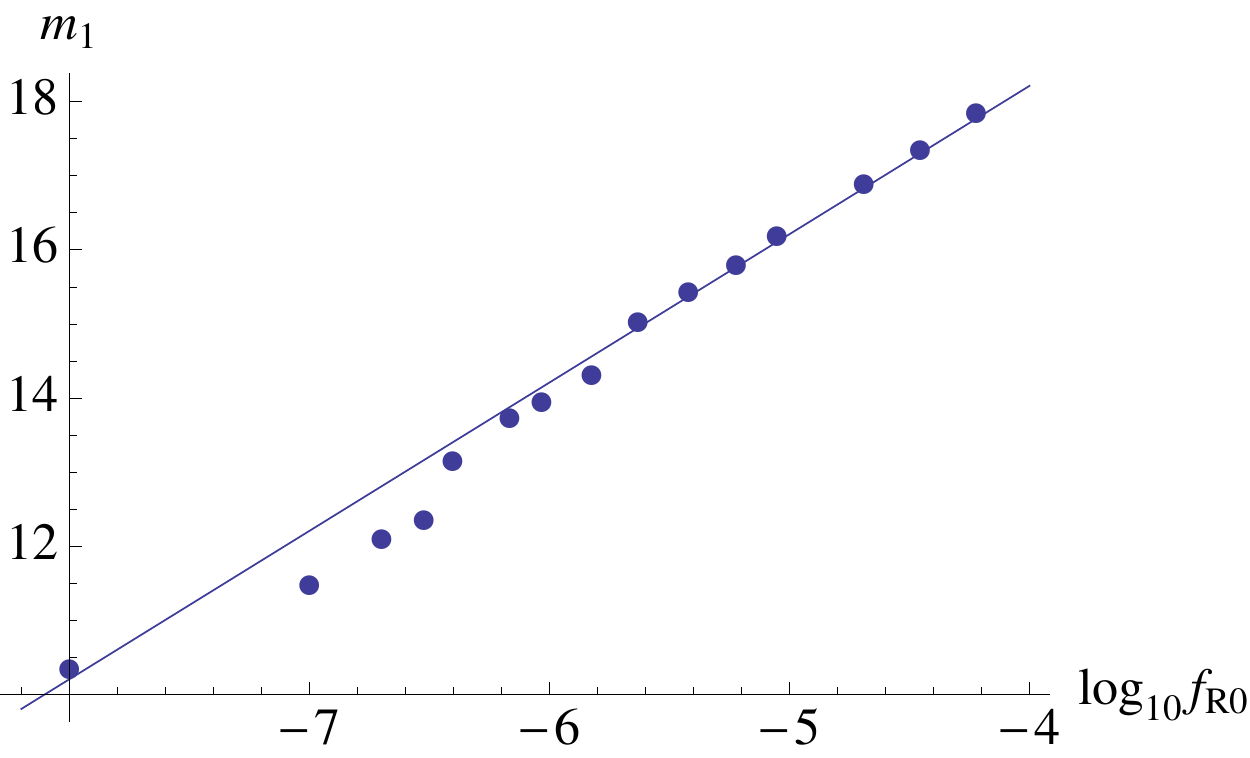}
\caption{Fit parameter $m_1$ ({\it dots}) as function of $f_{\rm R0}$ suggests a linear relation $m_1 = c_1 \log_{10}f_{\rm R0}+c_2$. The combined fit with 9 other fit parameters contained in the definitions of $b_2,b_3,a_3,a_4$ gives $c_1= 1.99$ and $c_2= 26.2$ ({\it full line}).}
\label{fig:m0ffr0}
\end{figure}

To quantify the effect of modified gravity we provide an interpolation function to fit the data. From Fig.\,\eqref{fig:deltacdetaandfit} we can impose the following ansatz for $\delta_{c}$

\begin{align} \label{deltacfit}
\delta_c(z,M,f_{\rm R0})&= \nonumber
\delta^{\Lambdaup}_c(z) \Bigg\{ 1+ b_2 (1+z)^{-a_3} \left( m_{b} -\sqrt{m_{b}^2+1}\right)+\\
& \qquad \qquad \ + b_3(\tanh m_{b}-1) \Bigg\}\\
m_{b}(z,M,f_{\rm R0})&=(1+z)^{a_3}\left(\log_{10} [M/(M_\odot h^{-1})]-m_1(1+z)^{-a_4}\right) \nonumber\\
m_1(f_{\rm R0}) &=  1.99 \log_{10}f_{\rm R0}+26.21 \nonumber\\
b_2 &= 0.0166\nonumber\\
b_3 (f_{\rm R0}) &=0.0027 \cdot (2.41-\log_{10}f_{\rm R0} ) \nonumber\\
a_3(f_{\rm R0}) &= 1 + 0.99 \exp\left[-2.08 (\log_{10}f_{\rm R0} + 5.57)^2\right]\nonumber\\
a_4(f_{\rm R0}) &= \left(\tanh\left[0.69\cdot (\log_{10}f_{\rm R0} + 6.65) \right] + 1\right) 0.11\nonumber
\end{align}
The fit function converges separately for $M\rightarrow \infty$ and $z\rightarrow \infty$ to its GR limit $\delta^{\Lambdaup}_c(z)$, which can be approximated by \cite{NS97}
\begin{equation}\label{deltacGR}
\delta^{\Lambdaup}_c(z) \simeq \frac{3(12 \pi)^{2/3}}{20}  \left(1 - 0.0123 \log_{10}\left[1 + \frac{\Omega_m^{-1} - 1}{(1 + z)^{3}}\right]\right)\,.
\end{equation}
We obtained our result \eqref{deltacfit} by considering  ${a_3,a_4,b_2,b_3,m_1}$ as independent fit parameters for each $f_{\rm R0}$ value.
For instance the various best fit parameters $m_1$ suggest a linear dependence  on $\log_{10}f_{\rm R0}$, see Fig.\,\ref{fig:m0ffr0}. In a similar fashion we obtain the other functional forms of $b_2,b_3,a_3,a_4$. The parameter $m_1$  is of particular interest since it determines the position of the chameleon transition at $z=0$, where $\delta_c(M)$ changes its behavior from a linear growth in $\log M$ to a constant; see Fig.\,\eqref{fig:deltacdetaandfit}. Therefore roughly speaking the halo mass function at $z=0$ approaches  $\Lambdaup$CDM for masses larger than
\begin{equation}\label{Mbreak}
M_1 = 10^{14.2}\left( \frac{f_{\rm R0}}{10^{-6}}\right)^2 M_\odot h^{-1}
\end{equation}
due to the chameleon mechanism.

\section{Halo Mass function: prediction for $\mathbold{f(R)}$ gravity and deviation from GR}
\label{section:massfunction}

\begin{figure}[t] 
\centering
\includegraphics[width=0.49 \textwidth]{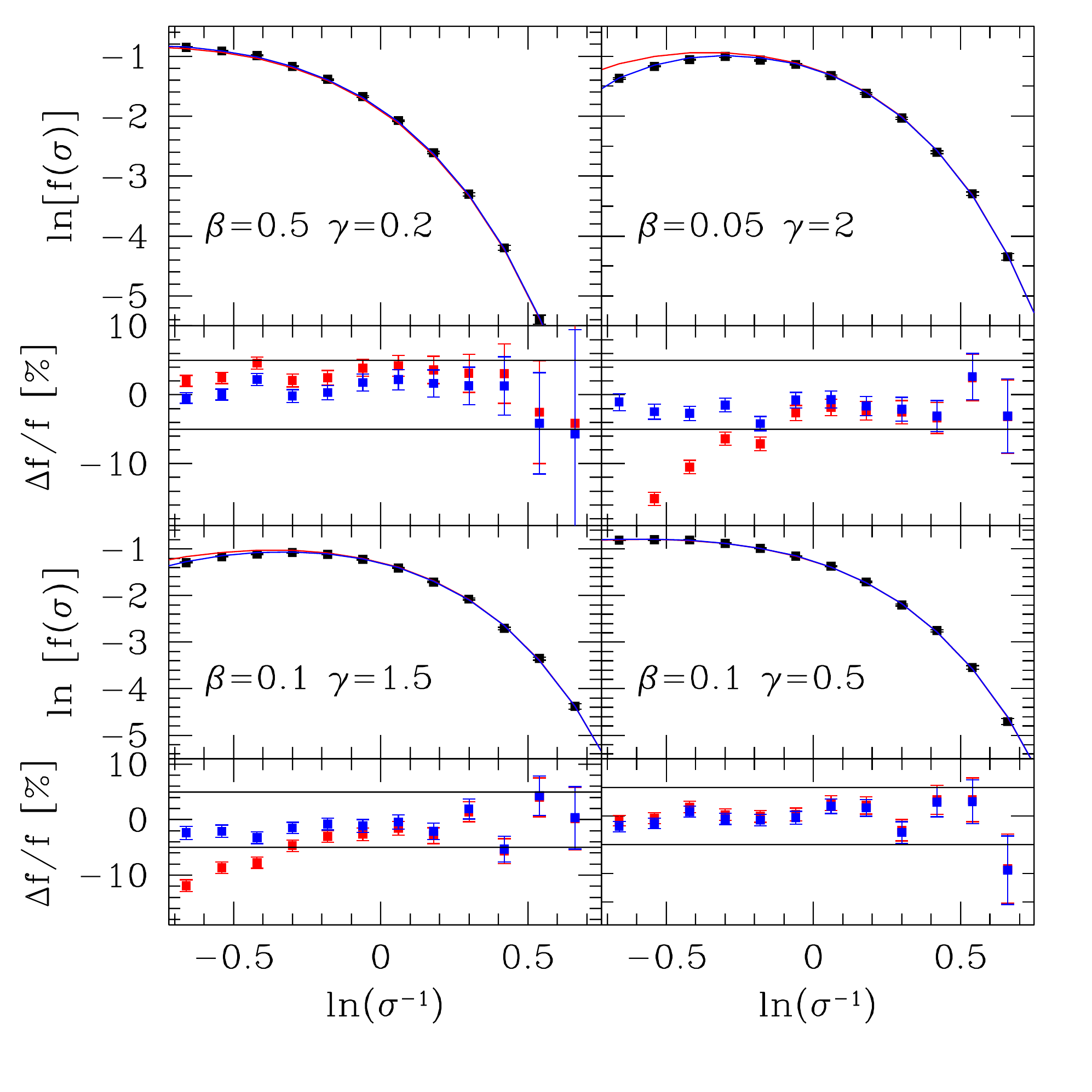}
\caption{Monte Carlo solution for different barrier compared to our formula Eq.\,\eqref{fskall} in blue and the one proposed by \cite{ST,DAMICO} of Eq.\,(\ref{fST}) in red. Relative differences are shown on the bottom panel.}
\label{hmfFig1}
\end{figure}

\begin{figure*}[t] 
\centering
\includegraphics[width=0.99 \textwidth]{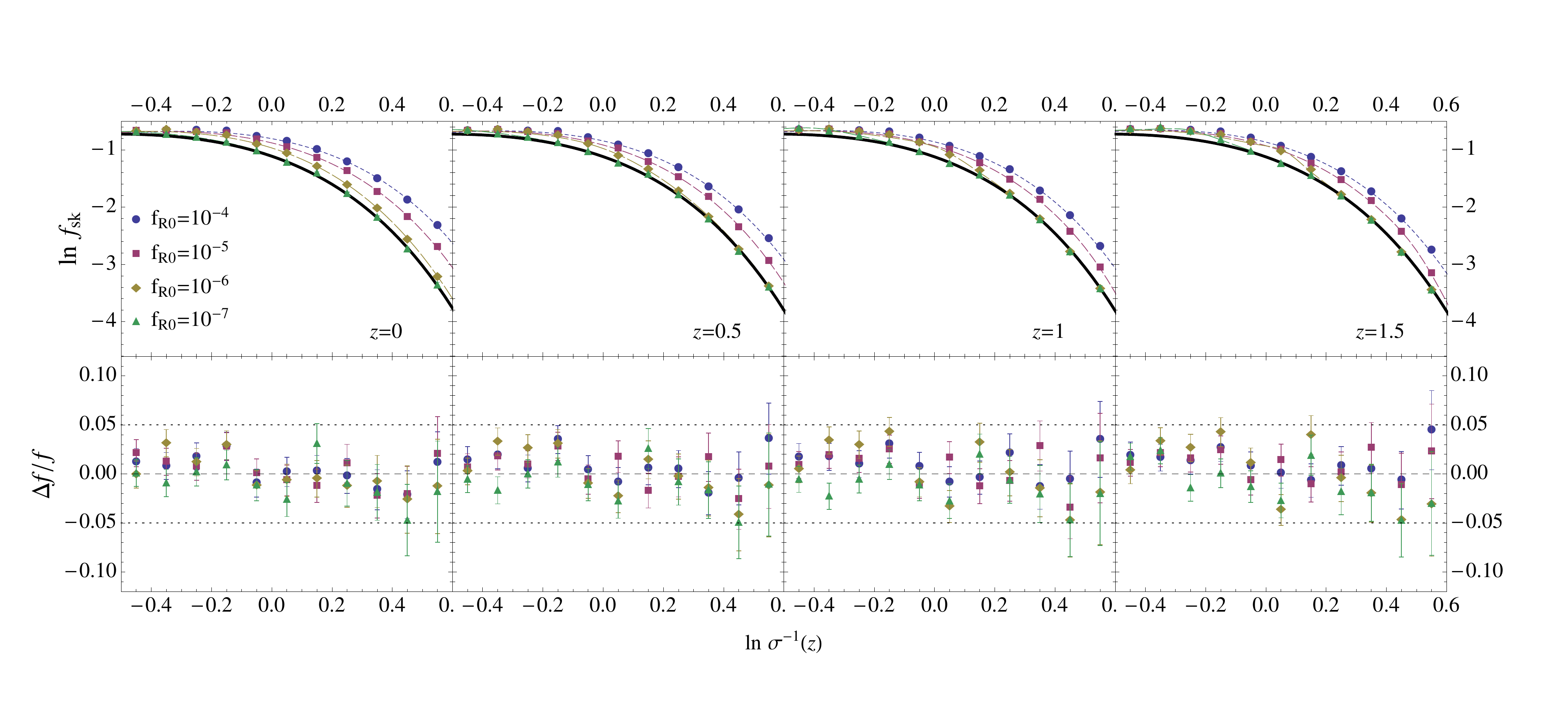}
\caption{Monte Carlo (dot) and theory (line) prediction for the $f(R)$ halo mass function at different redshift using spherical collapse barrier and sharp-$k$ filter on the upper panel. Black line show the GR prediction while the colours line are for different $f_{\rm R0}$. Lower panel shows the relative difference between the exact Monte Carlo solution and theory.}
\label{hmfFig2}
\end{figure*}

\begin{figure*}[t] 
\centering
\includegraphics[width=0.99 \textwidth]{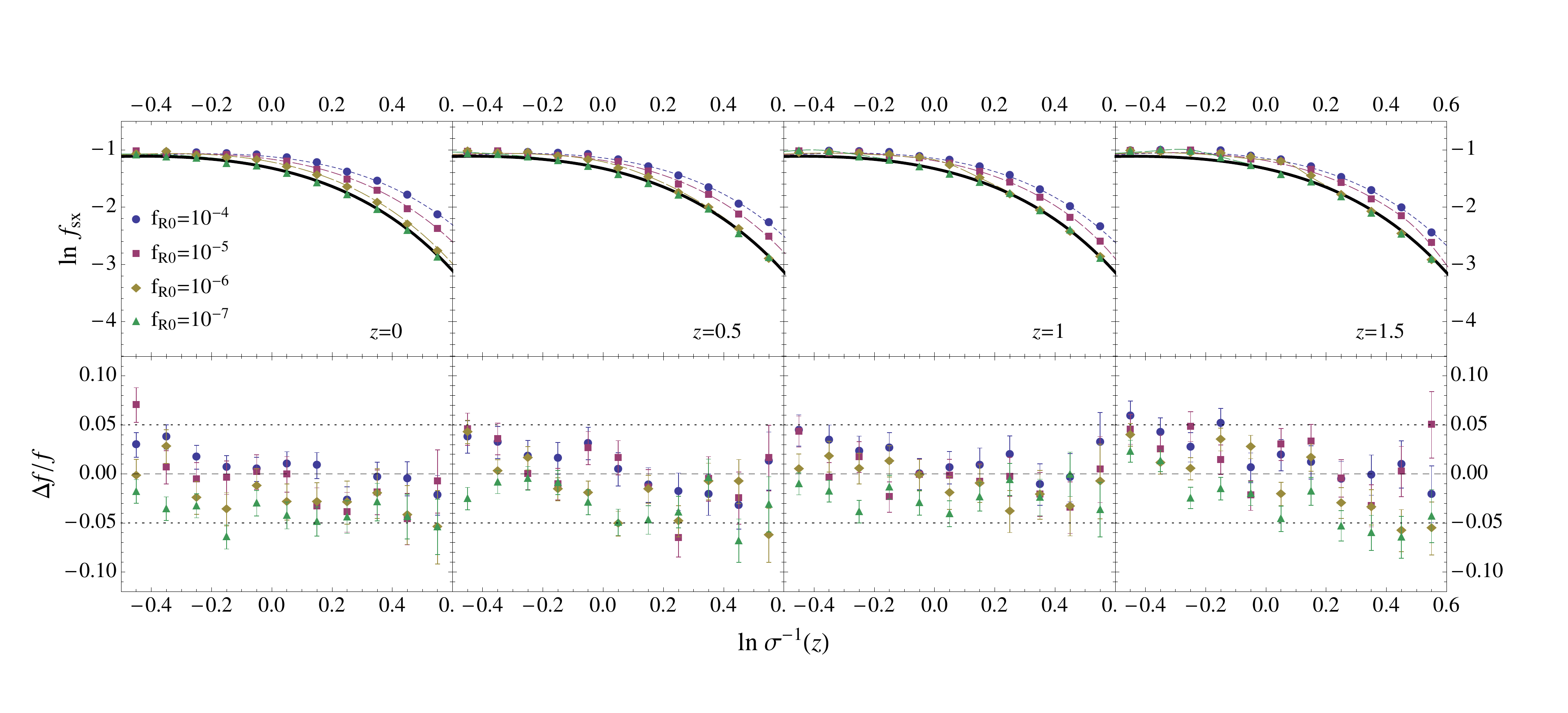}
\caption{Monte Carlo (dot) and theory (line) prediction for the $f(R)$ halo mass function at different redshift using a drifting diffusing barrier and sharp-$x$ filter on the upper panel. Black line show the GR prediction while the colours line are for different $f_{\rm R0}$. Lower panel shows the relative difference between the exact Monte Carlo solution and theory.}
\label{hmfFig3}
\end{figure*}

\begin{figure*}[t] 
\centering
\includegraphics[width=0.49 \textwidth]{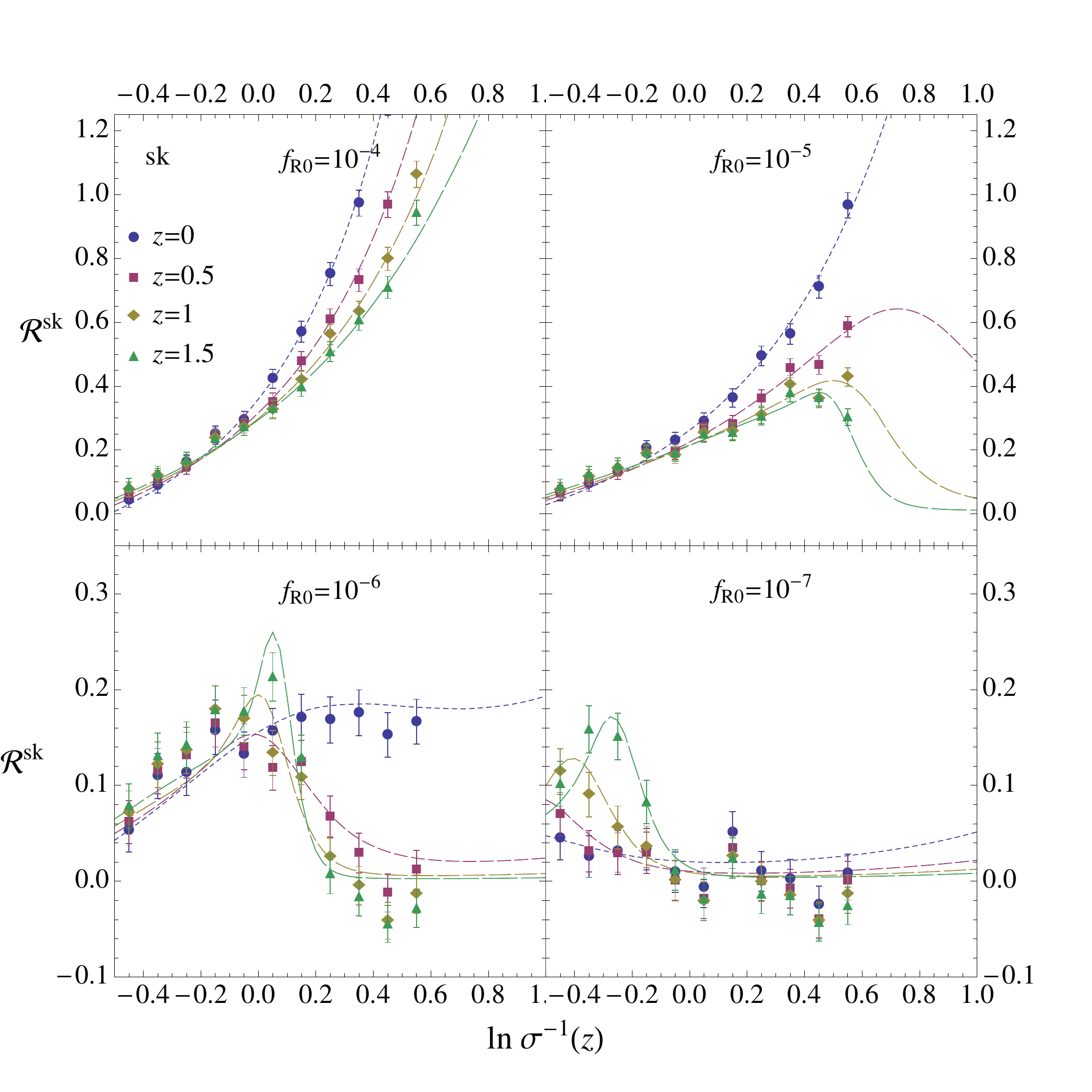}
\includegraphics[width=0.49 \textwidth]{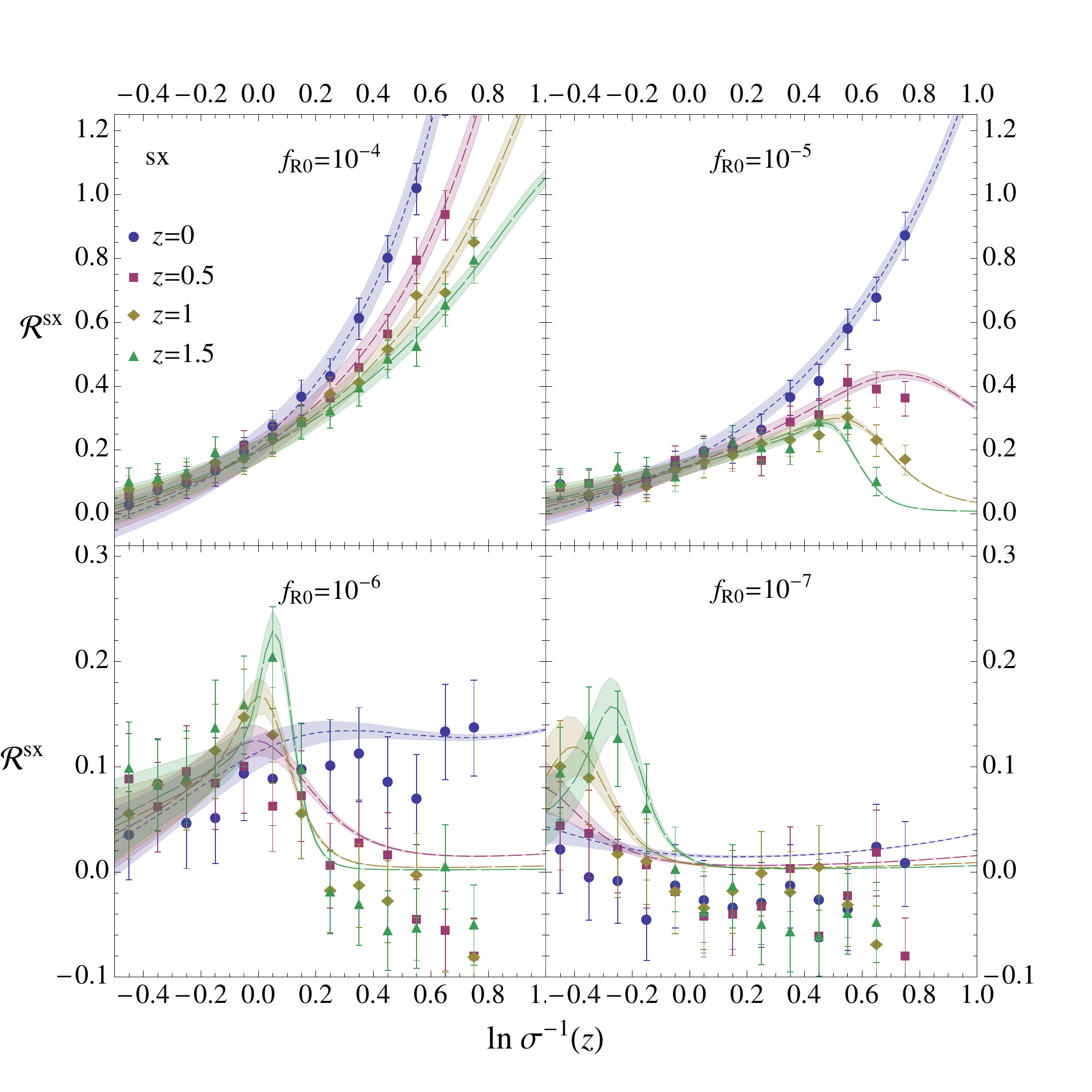}
\caption{Multiplicity function ratio $\mathcal{R}$ between GR and $f(R)$ gravity over different redshift and $f_{R0}$ parameters for naive spherical collapse with sharp-$k$ filter (left panel), see Eq.\,\eqref{variNODbandbeta} and using a drifting diffusive barrier with sharp-$x$ filter (right panel), see Eq.\,\eqref{variDbandbeta}. Colored bands in the rights panel show variations in $D_B$ and $\beta$, see Eq.\,\eqref{variDbandbeta}. Dots show the Monte Carlo run for $f(R)$ gravity.}. \label{hmfFig4}
\end{figure*}

Dark matter halos result from the non-linear collapse of initial density perturbations. The abundance of these virialized structures depends on both the properties of the initial matter density field and the collapse threshold which leads to their formation. Following the seminal work of \cite{PS74}, the excursion set approach \cite{BCEK91} computes the abundance of dark matter halos as a function of their mass. The method involves smoothing the initial density field over different realisations and positing that once the overdensity encapsulated in a smoothing region is above a threshold criteria, the region will collapse. The key assumption is then to equate the fraction of collapsed comoving volume to the comoving density of halos $n(M)$. Thus the number density of haloes in the mass range $[M, M+dM]$, the halo mass function $n(M)$, is given by 

\begin{equation}
n(M)=f(\sigma) \frac{\bar{\rho}_0}{M^2} \frac{d \ln \sigma^{-1}}{d \ln M}\,,\label{nandf}
\end{equation}
where $\bar \rho_0$ is the comoving background dark matter density and $f(\sigma)$ is related to fraction of collapsed volume. The fundamental quantity, which contains all information on the non-linear collapse dynamics, is $f(\sigma)$. In what follows we first review the analytic derivation of $f(\sigma)$ in case of spherical GR collapse. We then extend this calculation to $f(R)$ models with realistic collapse parameters. Having constructed the multiplicity function $f(\sigma)$ for these modified gravity models, we can provide an estimate of $f(R)$ signatures in the cluster abundance. Note that our methodology is different to existing approaches in the literature \cite{LL12,LL12b,LLKZ13}. In this work $f(R)$ effects are taken into account by averaging the barrier over environments of the initial Lagrangian perturbations.

\subsection{ Halo mass function prediction for uncorrelated random walk and generic barrier}\label{subsk}
To estimate the fraction of collapsed volume, one has to compute the probability $\Pi(R,\delta)$ of having an overdensity $\delta$  smoothed on a scale $R$. In the original Press-Schechter (PS) approach \cite{PS74}, assuming Gaussian initial conditions, the fraction of collapsed regions can be calculated analytically; it is given by 
\begin{equation}
F(R)=\int_{B}^{\infty} \Pi(\delta,\sigma(R)) d\delta\,,
\end{equation}
where $B$ is the collapse threshold and the probability density function (PDF) is  $\Pi(\delta,\sigma(R))=e^{-\delta^2/(2\sigma^2)}/\sqrt{2 \pi \sigma^2}$. However the PS approach suffers from the so called cloud in cloud problem: it requires an ad-hoc normalization of the mass function due to an incorrect counting of collapsed regions. To understand where the problem occurs let us review the standard excursion set procedure. We start by re-writing the smoothed overdensity on a scale $R$ at any random position as 

\begin{equation}
\delta(R)=\frac{1}{(2\pi)^3}\int d^3k\; W(k,R)\;\tilde{\delta}(z, k)\,,\label{deltasmooth}
\end{equation}
where $W$ and $\tilde{\delta}$  are the Fourier transforms of the filter function and and the linearly extrapolated $\delta$ respectively. Since $ \delta(R)$ is a random quantity, it was shown in \cite{BCEK91} that its evolution follows a Langevin equation. Once we fix the filter, there is a one to one relationship between the smoothing scale $R$, the mass of the halos $M(R)$ and the variance defined as 
\begin{equation}
S\equiv \sigma^2(z,R)=\frac{1}{2\pi^2}\int dk\;k^2 P(z,k)\;W^2(k,R). \label{Sdef}
\end{equation}
In the case of a sharp-$k$ filter $W(k,R)=\theta(1/R-k)$ and Gaussian initial conditions, the Langevin equation takes the form
\begin{equation}
\frac{\partial \delta}{\partial S}=\eta_\delta(S)\,,\label{Langevineq}
\end{equation}
where $\eta_\delta$ is white Gaussian noise completely specified by its mean $\langle \eta_\delta \rangle=0$ and variance $\langle \eta_\delta(S)\eta_\delta(S')\rangle=\delta_D(S-S')$. According to these equations, $\delta(R)$ performs a random walk and its evolution between two scales $S$ and $S'$ is determined by its previous step only. Since the system does not keep memory of previous steps, the dynamics corresponds to a Markovian random walk and the PDF follows a simple Fokker-Planck equation
\begin{equation}
\frac{\partial \Pi}{\partial S}=\frac{1}{2}\frac{\partial^2\Pi}{\partial \delta^2}\,.\label{FoPl}
\end{equation}

The PDF is fully specified by two initial conditions. At $S=0$, which corresponds to very large scales, the homogeneity of the universe implies that $\Pi(\delta,S=0)=\delta_D(\delta)$. If one relaxes the second condition then the solution of Eq.\,(\ref{FoPl}) is a Gaussian PDF corresponding to the original PS prediction. 

In the Excursion set approach, when random walks cross the threshold collapse $B$ at scale $S$ a halo of mass $M(S)$ is assumed to form. However random walks can cross $B$ more than once at different smoothing scales, and this can lead to double counting of halos. To evade this problem, one must remove walks when they cross $B$ for the first time. This can be encoded in an absorbing boundary condition; the PDF of uncollapsed objects is given by the solution of Eq.\,(\ref{FoPl}) with the second initial condition $\Pi(\delta\!\!=\!\!B,S)=0$. An exact analytic solution for a barrier that is a generic function of the smoothing scale does not exist. However for a constant spherical collapse barrier the exact solution is given by \cite{BCEK91,Redner}:
\begin{equation}
\Pi(\delta,S)=\frac{1}{\sqrt{2\pi S}}\left(e^{-\delta^2/(2S)}-e^{-(2\delta_{c}-\delta)^2/(2S)}\right)\,,
\end{equation}
\noindent where the first term on the right hand side is the previous Gaussian solution while the second term is known as the `anti-Gaussian'. The  fraction of collapsed volume is then
\begin{equation}
F(S)=1-\int_{-\infty}^{\delta_c}\Pi(\delta,S).
\end{equation}
The first-crossing rate is given by $\mathcal{F}(S)=dF(S)/dS$. From the definition of the multiplicity function $f(\sigma)=2\sigma^2\mathcal{F}(\sigma^2)$ it follows
\begin{equation}
f(\sigma)=\sqrt{\frac{2}{\pi}} e^{-\delta^{2}_c/(2\sigma^2)}\frac{\delta_c}{\sigma}\,,\label{fskGR}
\end{equation}
which is the original PS prediction with the correct normalisation. 

The above calculation corresponds to spherically collapsing overdensities; the situation is considerably more complicated in the real Universe. The dynamics of collapse is aspherical and small over-dense regions require additional matter to collapse \cite{SMT} since they are significantly affected by the surrounding shear field. Using ellipsoidal collapse in the excursion set approach introduces a stochastic barrier; this motivates the study of a generic barrier. In the $\Lambdaup$CDM case a simple Gaussian distribution for the barrier $B$ with a mean value $\bar B$ which drifts linearly as function of the variance $S$ is sufficient to reproduce the N-body halo mass function with high accuracy \cite{CA1,CA2,AC1,AC2}. Furthermore this barrier is consistent with the overdensity required to collapse measured in the initial condition \cite{achitouvetal} and has the advantage of admitting an exact solution for Markovian multiplicity function.

For $f(R)$ gravity we have shown in Section \ref{section:deltac} that spherical collapse cannot be modeled using a linear barrier. To obtain an analytical prediction for $f(\sigma)$ using a generic barrier, we start by introducing the variable $Y=B-\delta$ and  assume that the barrier is described by a Gaussian PDF with mean value $\bar{B}(S)$ and variance $D_B S$, with constant $D_B$. In such a scenario the Fokker-Planck equation for the $Y$ variable is given by

\begin{equation}
\frac{\partial \Pi(Y,S)}{\partial S}= \frac{1+D_B}{2}\frac{\partial^2\Pi(Y,S)}{\partial Y^2} -\frac{d\bar{B}}{dS} \frac{\partial \Pi(Y,S)}{\partial Y}\label{FPall}
\end{equation}

\noindent In the special case where $\bar{B}=\delta_c+\beta S$, the exact solution for $\Pi(Y,\delta)$ is \cite{CA1,CA2}
\begin{equation}
f(\sigma)=\sqrt{\frac{2a}{\pi}} e^{-a\bar{B}^2/(2\sigma^2)}\frac{\delta_c}{\sigma}
\end{equation}
with $a=1/(1+D_B)$. For generic $\bar{B}(S)$, the solution of the Fokker-Planck equation without the absorbing boundary condition is simply given by a Gaussian with mean $\bar{B}$ and variance $(1+D_B)S$. The crossing rate in this case would be given by

\begin{equation}
\mathcal{F}(S)=-\frac{d}{dS}\int_{0}^{\infty} \sqrt{\frac{a}{2\pi S}}e^{-a(Y-\bar{B})^2/(2S)}\,,
\end{equation}
leading to 
\begin{equation}
f(\sigma)=\sqrt{\frac{2a}{\pi}} e^{-a\bar{B}^2/(2\sigma^2)}\frac{1}{2\sigma}\left(\bar{B}-2S\frac{d\bar{B}}{dS}\right)\,.
\end{equation}

However, this expression does not have the correct normalisation  since we did not solve the equations using an absorbing boundary condition. For constant barrier, one could correct this expression by multiplying by an ad-hoc factor two, however for a linear drift this would not be sufficient to recover the exact solution since there is no factor of two multiplying the first derivative of $\bar{B}$. In fact one can show that the factor of two in front of the first derivative of $\bar{B}$ cancels once we add the anti-Gaussian term \cite{CA2}. Thus the exact solution for a constant and linear barrier is given by
\begin{equation}
f(\sigma)=\sqrt{\frac{2a}{\pi}} e^{-a\bar{B}^2/(2\sigma^2)}\frac{1}{\sigma}\left(\bar{B}-S\frac{d\bar{B}}{dS}\right).\label{fsktwoorder}
\end{equation}

\noindent Similarly for a generic barrier one could approximate the exact solution by expanding in higher order derivatives of the barrier and dividing these terms by a factor of two. Thus we propose the following formula for a generic barrier

\begin{equation}
\begin{split}
f(\sigma)&=\sqrt{\frac{2a}{\pi}} e^{-a\bar{B}^2/(2\sigma^2)}\frac{1}{\sigma}\Bigg(\bar{B}-\sigma^2\frac{d\bar{B}}{d\sigma^2} +\\
&\qquad\qquad\qquad\qquad+\frac{1}{2} \sum_{n\geq 2} \frac{(-\sigma^2)^n}{n!} \frac{d \bar{B}^n}{d(\sigma^2)^n} \Bigg)\,.\label{fskall}
\end{split}
\end{equation}

\noindent Note that this expression matches the first two terms of \cite{DAMICO} which are equivalent to \cite{ST} for $D_B=0$. In  \cite{ST} it was proposed that the additional correction for a generic barrier is given by


\begin{equation}
f_{\rm ST}(\sigma)=\sqrt{\frac{2}{\pi}} e^{-\bar{B}^2/(2\sigma^2)}\frac{1}{\sigma}\left(\sum_{n=0}^5 \frac{(-\sigma^2)^n}{n!} \frac{d \bar{B}^n}{d(\sigma^2)^n} \right)\,,\label{fST}
\end{equation}
see also \cite{dSMR11}.

\noindent Note that  (\ref{fskall}) or (\ref{fST}) are  approximations, and should only be applied to  barriers well-described by algebraic functions for which all derivatives are well-defined. In order to test the robustness of our ansatz, we performed a series of Monte Carlo random walks for various barrier models   following the procedure described in \cite{BCEK91}. In the case where $D_B=0$ and $\bar{B}=\delta_c+\beta S^\gamma$, we exhibit some of the Monte Carlo random walks in Fig.\,\ref{hmfFig1}. We find that for barriers which scale like $0<\gamma<1$ and $\beta<1$, the first order derivative term is sufficient to fit the Monte Carlo walks with high accuracy while for $\gamma>1$ and $\beta <1$ we require the fourth/fifth term to obtain an accurate match. In such a case, $f_{\rm ST}$ is unable to reproduce the exact solution while Eq.\,(\ref{fskall}) fits these Monte Carlo random walks with high accuracy. In what follows we will use Eq.\,(\ref{fsktwoorder}) to model the spherical collapse barrier of $f(R)$ but the method that we apply in section \ref{sxsection} to predict the halo mass function can be extended to non-standard GR where the barrier can be an arbitrary algebraic function of the form $\sum_n \beta^{n} S^{n}$ with $\beta<1$ and $\gamma>0$. 


In the absence of N-body simulations, one way to evaluate the effect of $f(R)$ gravity on the halo mass function is to use spherical collapse (ie: $D_B=0, \beta=0$) and measure the ratio between the GR and $f(R)$ prediction for an uncorrelated walk (ie: sharp-$k$ filter). For that purpose we first need an accurate prediction for $f(R)$ gravity. We run Monte Carlo walks for various $f_{\rm R0}$ parameters to test the accuracy of Eq.\,(\ref{fsktwoorder}). 

In Fig (\ref{hmfFig2}) we observe the halo mass function corresponding to the exact Monte Carlo solution (dot) and our prediction (full line) Eq.\,(\ref{fsktwoorder}) with $\beta=0, D_B=0$ and $\delta_c$ given by Eq.\,\eqref{deltacfit}. On the lower panel we show the relative difference between the Monte Carlo and theoretical prediction. We see that the difference is of order $\sim 5\%$, confirming that Eq.\,(\ref{fsktwoorder}) provides an excellent fit. The colours correspond to the different model parameters we test: blue, red, yellow and green correspond to $\log_{10}f_{\rm R 0}=-4$, $\log_{10}f_{\rm R 0}=-5$, $\log_{10}f_{\rm R 0}=-6$ and $\log_{10}f_{\rm R 0}=-7$ respectively and black is the GR spherical collapse prediction. The deviation between GR and $f(R)$ gravity is explored further in section \ref{deviationGR}.

\subsection{Modelling of the halo mass function using realistic mass definition and collapse parameters}\label{sxsection}

In the previous section we predicted the $f(R)$ multiplicity function using two simple assumptions: one is related to the filtering procedure which we took to be sharp-$k$ and the other to the spherical dynamics of collapse. In fact, the Fokker-Planck equation (\ref{FPall}) is only valid in the special case where there is no absorbing boundary \cite{MR1} or if the random walk is Markovian. This is the case only when $\delta$ is smoothed with a sharp$-k$ filter in Eq.\,(\ref{deltasmooth}). The choice of filter is important as it defines the relationship between the mass of the halos and the variance of the field. Assuming that the mass of a halo is given by $M(R)=\bar{\rho}_0 V_{\rm sp}$ where $V_{\rm sp}(R)$ is the volume of a sphere, then one should actually consider a real-space top-hat filter (ie: sharp$-x$), where the Lagrangian radius of the halo is related to the variance $\sigma(R)^2$ in Eq.\,(\ref{Sdef}), which we normalise to $\sigma_8=0.8$. In this case there is no exact analytical solution for the PDF. 

In \cite{MR1} a path integral approach to compute the non-Markovian corrections induced by a sharp$-x$ filter has been developed.  The magnitude of the correction is given by $\kappa$, which depends on the linear matter power spectrum. For a standard $\Lambdaup$CDM Universe,  $\kappa\sim 0.65$. In \cite{MR2} this formalism was applied to a stochastic barrier with Gaussian distribution and in \cite{CA1,CA2} the solution was extended to a diffusive barrier with mean $\delta_c+\beta S$. Such a barrier encapsulates the main features of ellipsoidal collapse. In such a case, the multiplicity function to first order in $\kappa$ is given by 

\begin{equation}
f(\sigma)=f_0(\sigma)+f_{1,\beta=0}^{m-m}(\sigma)+
f_{1,\beta^{(1)}}^{m-m}(\sigma)+f_{1,\beta^{(2)}}^{m-m}(\sigma)\,,\label{ftot}
\end{equation}

where 
\begin{equation}
 f_0(\sigma)=\frac{\delta_c}{\sigma}\sqrt{\frac{2a}{\pi}}\,e^{-\frac{a}{2\sigma^2}(\delta_c+\beta\sigma^2)^2},\label{fsigma0}
\end{equation}

\begin{equation}
f_{1,\beta=0}^{m-m}(\sigma)=-\tilde{\kappa}\dfrac{\delta_c}{\sigma}\sqrt{\frac{2a}{\pi}}\left[e^{-\frac{a \delta_c^2}{2\sigma^2}}-\frac{1}{2} \Gamma\left(0,\frac{a\delta_C^2}{2\sigma^2}\right)\right]\,,\label{beta0}
\end{equation}
\begin{equation}
f_{1,\beta^{(1)}}^{m-m}(\sigma)=- a\,\delta_c\,\beta\left[\tilde{\kappa}\,\text{Erfc}\left( \delta_c\sqrt{\frac{a}{2\sigma^2}}\right)+ f_{1,\beta=0}^{m-m}(\sigma)\right]\,,\label{beta1}
\end{equation}
\begin{equation}
f_{1,\beta^{(2)}}^{m-m}(\sigma)=-a\,\beta\left[\frac{\beta}{2} \sigma^2 f_{1,\beta=0}^{m-m}(\sigma)+\delta_c \,f_{1,\beta^{(1)}}^{m-m}(\sigma)\right]\,.\label{beta2}
\end{equation}

In \cite{achitouvetal} it was shown that the first order approximation in $\kappa$ is sufficient to reproduce the exact solution to $\sim 5\%$ accuracy, using parameter values $\beta=0.12$, $D_B=0.4$. This effective barrier can match the N-body halo mass function with accuracy $\sim 5\%$ and is also consistent with the collapse threshold  measured in the initial conditions, suggesting that $\beta,D_B$ are parameters that should depend on physics of the collapse dynamics. For $f(R)$ spherical collapse we find that we recover the general relativistic prediction for massive halos, however for small mass objects the threshold of collapse decreases as function of the variance. Such behaviour can be roughly approximated by a negative drift coefficient $\beta$ which would counteract the expected $\beta >0$ behaviour associated with GR ellipsoidal collapse. Thus it is not clear how $f(R)$ gravity will effect the collapse of an aspherical patch. However, we can reasonably assume that for $f_{\rm R0} \rightarrow 0$ one should recover the GR limits. As an initial step we can therefore fix $\beta$ and $D_B$ to their GR values and run MC walks for the sharp-$x$ filter, with $\delta_c$ given by \eqref{deltacfit}. In what follows we exhibit the resulting halo mass functions, and end the section by estimating the sensitivity of our results to our choice of $(\beta,D_B)$.\\

In order to predict the multiplicity function for $f(R)$ gravity we begin by noting that the sharp-$x$ multiplicity function can be rewritten as the sharp-$k$ function with a correction in $\kappa$. Hence the ratio between the GR and $f(R)$ predictions is given by 

\begin{equation}
\frac{f^{f(R),\rm sx}}{f^{\rm GR,sx}}=\frac{f^{f(R),\rm sk}+ f^{f(R)}_{\kappa=1}+\theta(\kappa^2) }{f^{\rm GR,sk} +f^{\rm GR}_{\kappa=1}+\theta(\kappa^2)}\,.
\end{equation}
Therefore 
\begin{equation}
f^{f(R),\rm sx}(\sigma)= \frac{f^{\rm GR,sx}}{f^{\rm GR,sk}} \left[f^{f(R),\rm sk}+ \left(f^{f(R)}_{\kappa=1}-f^{\rm GR}_{\kappa=1}\right) +\theta(\kappa^2)\right]\,,
\end{equation}
\noindent where $f^{\rm GR,sk}$ is given by Eq.\,(\ref{fsigma0}), $f^{\rm GR,sx}$ by Eq.\,(\ref{ftot}), $f^{f(R),\rm sk}$ by Eq.\,(\ref{fsktwoorder}), $f^{\rm GR}_{\kappa=1}$ by Eq.\,(\ref{beta0}-\ref{beta2}) and $f^{f(R)}_{\kappa=1}$ is the first order non-Markovian corrections due to the sharp-$x$ filter. Since this correction is also proportional to $\kappa$, it seems reasonable that the difference $f^{f(R)}_{\kappa=1}-f^{\rm GR}_{\kappa=1}$ should be negligible. In fact one could rewrite $f^{f(R)}_{\kappa=1}$ as an expansion around the GR spherical collapse solution, in which case the first order term would be given by Eq.\,(\ref{beta0}) as for the GR case (ie: $\beta=0$) and the first non-vanishing term in the difference would be proportional to $a\,\kappa\,\beta\sim 0.05$. Thus we assume in what follows that 
\begin{equation}
f^{f(R),\rm sx}(\sigma)\simeq f^{\rm GR,sx}(\sigma) \frac{f^{f(R),\rm sk}}{f^{\rm GR,sk}}\,. \label{sxprediction}
\end{equation}

We test Eq.\,(\ref{sxprediction}) by comparing with the exact Monte Carlo solution. The result is shown in Fig.\,(\ref{hmfFig3}), where we use the GR parameters for $(\beta,D_B)$. In the upper panel we see the Monte Carlo solution (dots) and Eq.\,(\ref{sxprediction}) for different redshift and $f_{\rm R0} $ parameters. On the bottom panel we show the relative difference between Monte Carlo runs and equation (\ref{sxprediction}). Once again, the fractional difference is of order $\sim 5\%$ confirming the validity of Eq.\,(\ref{sxprediction}). Hence we adopt this simple prescription to define the multiplicity function for $f(R)$ gravity. The halo mass function can obtained from Eq.\,\eqref{sxprediction} via
\begin{equation}
n(M, z,f_{\rm R0})=f^{f(R),\rm sx}(\sigma) \frac{\bar{\rho}_0}{M^2} \frac{d \ln \sigma^{-1}}{d \ln M}\label{nandfofsigma}\,,
\end{equation}
where  $\sigma(z,R)= D(z)\, \sigma(z\!\!=\!\!0,R)$ is calculated from $\Lambdaup$CDM linear growth $D(z)$ and the linear power spectrum $P(z\!\!=\!\!0,k)$ obtained from CAMB as described in Sec.\,\ref{subsec:Sphericalcollapse}.

\subsection{Realistic prediction for $f(R)$ gravity and deviation from GR}\label{deviationGR}

For completeness we test whether there is a significant modified gravity imprint on the $f(R)$ mass function Eq.\,\eqref{nandfofsigma}. To evaluate the sensitivity of our results to our choice of collapse parameters, we consider the ratio of the $f(R)$ and GR predictions using different values of $(\beta,D_B)$. One might reasonably assume that if the spherical collapse deviation with respect to GR is $$\eta \equiv(\delta_{c}^{f(R)}-\delta_{c}^{\rm GR})/\delta_{c}^{\rm GR}$$ then the drift term $\beta$ should also exhibit deviations of order $\eta$. Since the diffusive term $D_B$ appears as $\delta_c/\sqrt{2S(1+D_B)}$ in the mass function, we estimate that $\sqrt{D_B}$ will also be modified on the order of $\eta$. We therefore define 
\begin{equation} \label{variDbandbeta}
\begin{split}
&\mathcal{R}^{\rm sx}=\frac{f^{f(R),\rm sx}(\delta_{c}^{f(R)},\beta^{\rm GR},D_{B}^{\rm GR})}{f^{\rm GR,sx}}-1\\
&\mathcal{R}^{\rm sx}_+=\frac{f^{f(R),\rm sx}(\delta_{c}^{f(R)},\beta^{+},D_{B}^{+})}{f^{\rm GR,sx}}-1\\
&\mathcal{R}^{\rm sx}_-=\frac{f^{f(R),\rm sx}(\delta_{c}^{f(R)},\beta^{-},D_{B}^{-})}{f^{\rm GR,sx}}-1\\
\end{split} 
\end{equation}

\noindent where $f^{f(R),\rm sx}$ is given by Eq.\,(\ref{sxprediction}), $f^{\rm GR,sx}$ is given by Eq.\,(\ref{ftot}) with $D_{B}^{\rm GR}=0.4, \beta^{\rm GR}=0.12$ and $\beta^{\pm}\equiv\beta^{\rm GR}(1\pm  \eta)$ while $D_{B}^{\pm}\equiv D_{B}^{\rm GR} (1 \pm\eta^2)$.


In left panel of Fig.(\ref{hmfFig4}) we exhibit the ratios between the $f(R)$ and GR halo mass functions using the naive sharp-$k$ and spherical collapse for four model parameters using Eq.\,(\ref{fsktwoorder}) in section \ref{subsk}:
\begin{equation}\label{variNODbandbeta}
\mathcal{R}^{\rm sk}\equiv\frac{f^{f(R),\rm sk}(\delta_{c}^{f(R)},\beta=0,D_B=0)}{f^{\rm GR,sk}}-1.
\end{equation}

\noindent On the right panel we show the ratio $\mathcal{R}^{\rm sx}$ as lines and $\mathcal{R}^{\rm sx}_+,\mathcal{R}^{\rm sx}_-$ as shaded strips. First note that for a given scale $S$, $\mathcal{R}^{\rm sx}$ and $\mathcal{R}^{\rm sk}$ differ significantly, implying that a simple spherical collapse model in the excursion set framework with a sharp-$k$ filter should not be used to measure departures from GR. However both $\mathcal{R}^{\rm sx}$ and $\mathcal{R}^{\rm sk}$ share the same qualitative features. Our second important conclusion is that varying $D_B$ and $\beta$ over the range we might expect in $f(R)$ gravity does not significantly modify the departure from GR. Indeed, the width of the strip does not appreciably change for the multiplicity function. This serves as a final check of our analysis. \\

\begin{figure}[t] 
\centering
\includegraphics[width=0.49 \textwidth]{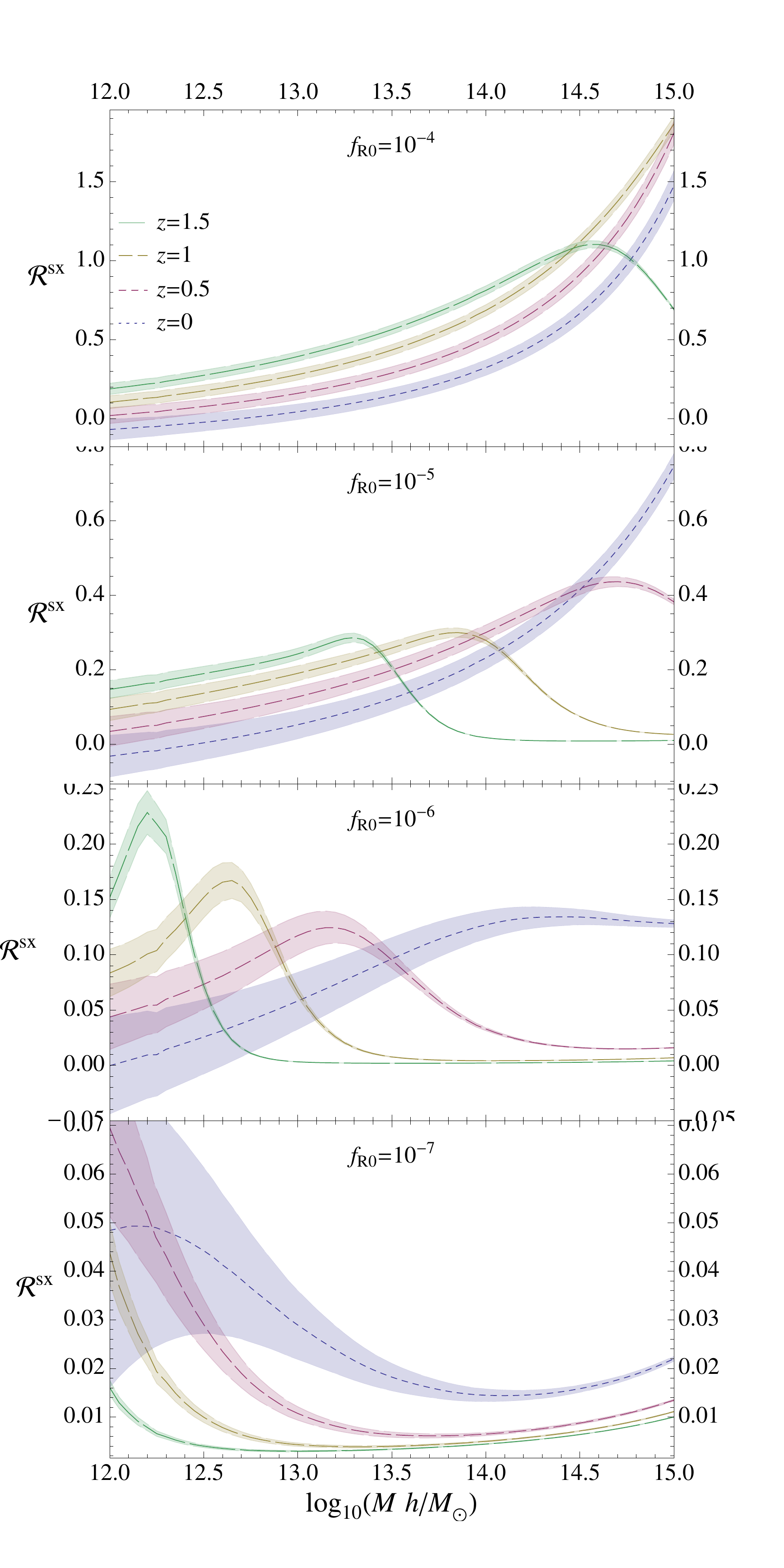}
\caption{The panels from top to bottom show the halo mass function ratios \eqref{variDbandbeta} for different $f_{\rm R0}$. Within each panel different lines show different collapse redshifts (see the legend in the first panel).}
\label{hmfFig5}
\end{figure}

\begin{figure}[t] 
\centering
\includegraphics[width=0.49 \textwidth]{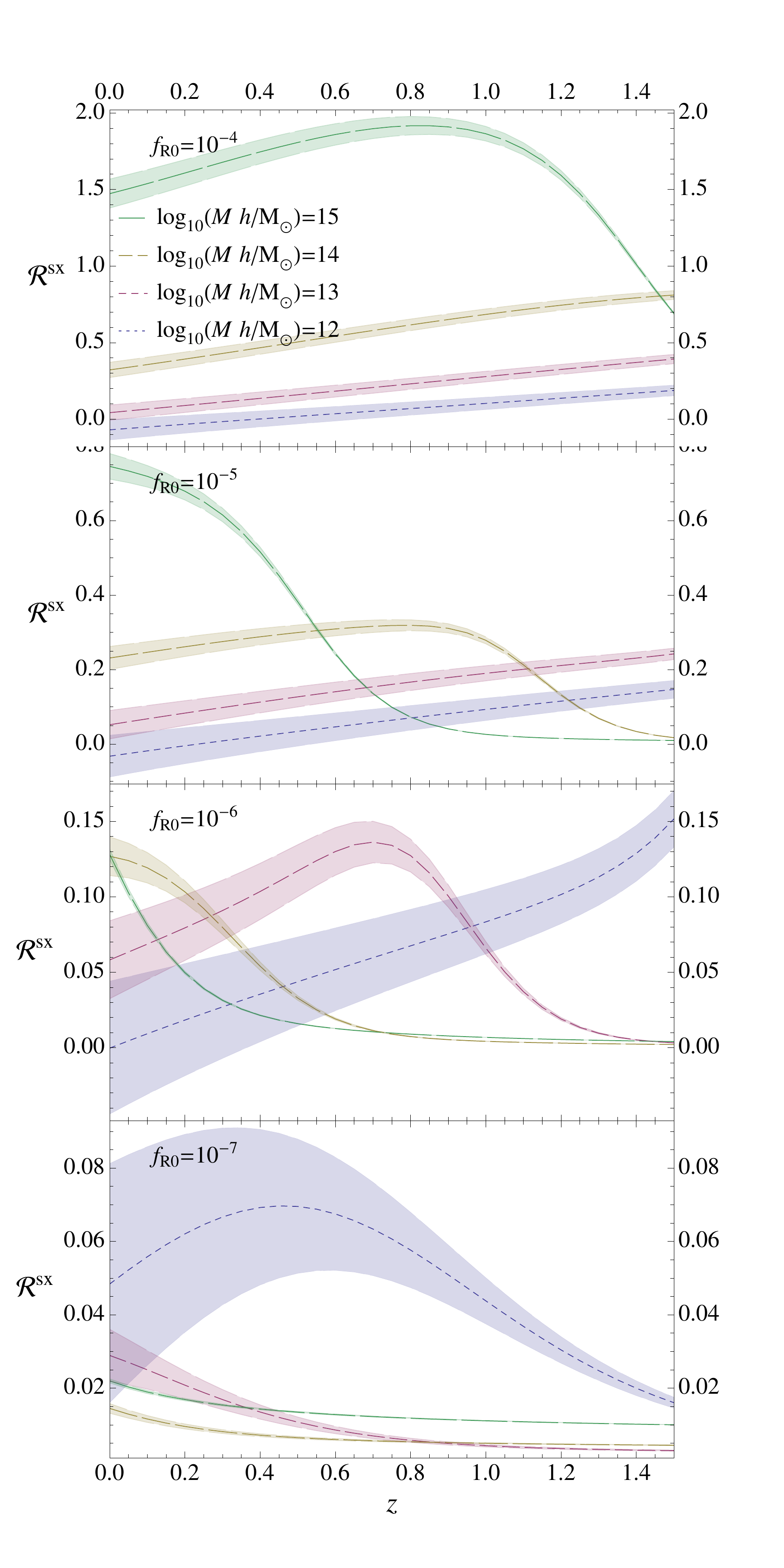}
\caption{The panels from top to bottom show the redshift evolution of the halo mass function ratio  \eqref{variDbandbeta} for different $f_{\rm R0}$. Within each panel the different lines show different halo masses (see the legend in the first panel).}
\label{hmfFig6}
\end{figure}

Finally, we use Eq.\,(\ref{sxprediction}) in Eq.\,(\ref{nandf}) to study how the number count of halos changes for $f(R)$ gravity compared to GR. For this we note that from Eq.\,\eqref{nandfofsigma} follows that
\begin{equation}
\mathcal{R}^{\rm sx}=\frac{n^{f(R),\rm sx}(\delta_{c}^{f(R)},\beta^{\rm GR},D_{B}^{\rm GR})}{n^{\rm GR,sx}}-1\,.
\end{equation}
 In Figs.\,\ref{hmfFig5} we show the number count ratio for various $f_{\rm R0}$ values and the evolution at different redshifts. It is clear from this figure that the $f(R)$ signature strongly depends on redshift. In Fig.\,\ref{hmfFig6}  we exhibit the redshift evolution of the $f(R)$-GR halo mass function ratio for various mass bins and $f_{\rm R0}$ values. There is a distinctive signature in both the mass and time dependence of the halo mass function due to the chameleon effect. The lower panels of Figs.\,(\ref{hmfFig5},\ref{hmfFig6}) correspond to field values $f_{\rm R0}=10^{-7}$. Modified gravity effects are suppressed for models so close to GR, suggesting a floor $f_{\rm R0} \sim O(10^{-7})$ below which  cluster counts will not competitively probe non-standard physics.  Improvements to our results could be made by testing the abundance of low mass halos by using N-body $f(R)$ simulations \cite{PBS13}. This will allow us to further study aspherical collapse and could ameliorate the uncertainties that exist in our numerical results (shown by the shaded regions in our figures).
 
As was shown in \cite{JVS12}  looking for local variations of physical properties induced by the environment-dependent chameleon effect puts the strongest constraints on $f_{\rm R0}$. In the context of spherical collapse, taking into account local variations would require the use of initial density profiles conditioned on the desired environment. For instance one could use as initial condition for spherical collapse the mean shape not only conditioned on the height and mass of the peak (see Eq.\,\eqref{peaksformula}), but also conditioned on value of the density at a relevant scale; the environmental density $\delta_{\rm env}$. 

The resulting conditional spherical collapse threshold $\delta_c(z,M,f_{\rm R0}, \delta_{\rm env})$ would then allow the construction of a conditional halo mass function \cite{LZK11,LE11, LL12, LL12b, LLKZ13}. This has not been done for the above mentioned physically motivated profile conditioned on $\delta_{\rm env}$.

\section{Conclusion}
\label{section:conclusion}

Confronting modified gravity models with cosmological data sets is a highly non-trivial task. Even a seemingly straightforward physical process such as the collapse of a spherically symmetric overdensity becomes a problem fraught with complications. In this work we have calculated numerically the linear threshold for collapse $\delta_{c}$ for one of the simplest modified gravity models in the literature; $f(R)$ gravity. By solving the full modified Einstein and fluid equations, we were able to construct an approximate functional form for $\delta_{c}(z_{c},M, f_{\rm R0})$, which  depends on both the initial size and shape of the overdensity and also the modified gravity parameter. A number of subtleties were encountered, such as the choice of initial conditions and the applicability of the linearization procedure. 

Using the spherical collapse $\delta_{c}$ of $f(R)$ gravity and a drifting diffusing barrier in the excursion set approach, we constructed a physically motivated halo mass function using the formalism first introduced in \cite{MR2, CA1}. This method has been shown to accurately reproduce the general relativistic halo mass function, and we expect that it is also robust for a wide variety of modified gravity models. It was shown in Section \ref{section:massfunction} that our ansatz for $n(M, z,f_{\rm R0})$ is in excellent agreement with our numerical Monte Carlo random walk simulations, and can be applied to generic barriers that are algebraic functions of the variance. 

Whilst the collapse threshold that we obtain is based upon the $f(R)$ spherical collapse barrier, we have gone beyond  simple spherical collapse when calculating $n(M,z,f_{R0})$. Our ansatz introduces two parameters;  $\beta$ takes into account deviations from spherical collapse  and $D_B$ quantifies the scatter around it. In addition we have shown that our results are relatively insensitive to possible deviations to these parameters induced by modified gravity. The existence of substructure in the halo progenitor environment \cite{LH11}, which is partially accounted for in our work, and substructure within the halo progenitor, see \cite{LE11}, influences the chameleon effect and further complicates the computation of the halo mass function. Therefore more work is required to fully understand aspherical collapse and all effects of modified gravity on the multiplicity function. The next step in this direction would be to directly compare our approach with modified gravity N-body simulations \cite{PBS13} and measure the parameters of collapse following \cite{achitouvetal}.

\label{section:comparison}
\acknowledgements 
We would like to thank Lucas Lombriser and Tsz Yan Lam for useful discussions, Eric Linder for comments on an early version of the draft and Ravi Sheth for correcting some details. We would also like to thank Marco Baldi and Ewald Puchwein for allowing us to confirm our mass function results using their N-body simulation mass function data. This work has been supported by World Class University grant R32-2009-000-10130-0 through the National Research Foundation, Ministry of Education, Science and Technology of Korea. I. Achitouv and J. Weller acknowledge
support from the Trans-Regional Collaborative Research
Center TRR 33 ``The Dark Universe'' of the
Deutsche Forschungsgemeinschaft (DFG).
\newpage

\appendix

\section{Nonlinear equations}
\label{appendix:nonlineareq}
We introduce the following notation. $K$ is the trace of the extrinsic curvature
\begin{equation}
K=3 e^{-\Phi}(\dot{\Psi}-H)\,,
\end{equation}
$H=\dot{a}/a$ is the Hubble constant of the asymptotic FRW and $\lapl$ is the flat space Laplacian. The general relativistic version of the $\gamma$-factor, $w$,  is given by
\begin{equation}
w=e^{\Phi} u^0=\frac{1}{\sqrt{1-v^2 e^{-2(\Phi+\Psi)}}}\,,\label{gammafactor}
\end{equation}
and $v\equiv v^r\equiv a u^r/u^0$ is the radial coordinate speed times $a$. The velocities $u_r$ and $v$ are related via
\begin{equation}
u_r=v w a e^{-2\Psi-\Phi}\,.
\end{equation}
The ``momentum''\footnote{This is not the canonical momentum of $\varphi$, see \cite{DSY09}} of $\varphi$ is denoted by
\begin{equation}
 \Pi\equiv e^{-\Phi} \dot{\varphi}\,.\label{defPi}
\end{equation}
\begin{widetext}
\noindent The Einstein equations can be decomposed into energy constraint ($-\ ^0_{\ 0}$)
\begin{subequations}\label{fofRequations}
\begin{equation}
 \frac{1}{3}K^2-K \Pi +e^{2 \Psi}a^{-2}\left(\lapl (2\Psi-\varphi)-\Psi'^2+\Psi'\varphi'-\varphi'^2\right)= e^{-\varphi}\left(\kappa^2\rho w^2+\frac{1}{2}V\right) \,,
\end{equation} 
momentum constraint ($e^{-\Phi}\cdot\ _{0 r}$)
\begin{equation}
 \frac{2}{3}K'-\left(\Pi'+\Pi \varphi'+\frac{1}{3}K \varphi' \right) = -e^{-\varphi} \kappa^2\rho w u_r\,,
 \end{equation}
evolution equation ($^j_{\ j}$)
 \begin{align}
 e^{-\Phi} (2 \dot{K}-3\dot{\Pi})-K^2+2K \Pi-3\Pi^2-2e^{2 \Psi} a^{-2} \lapl(\Psi-\Phi-\varphi)-\qquad \qquad \qquad\nonumber\\-2e^{2 \Psi} a^{-2}\left( -\Phi'^2-\frac{1}{2}\Psi'^2+\Phi'\Psi'-\varphi'^2+\Psi' \varphi' -\frac{3}{2}\Phi' \varphi' \right)= &-e^{-\varphi}\left(\kappa^2\rho (1-w^2)+\frac{3}{2}V\right)\,,\label{spattrace}
 \end{align}
 and Newtonian gauge condition ($_{rr}-1/3\ ^j_{\ j}$)
 \begin{equation}
 \lapl (\Psi-\Phi-\varphi)-\frac{3}{r}(\Psi'-\Phi'-\varphi')+\Psi'^2-\Phi'^2-\varphi'^2-2\Psi'\Phi'-2\Psi' \varphi'= 0\,.\label{spattraceless}
 \end{equation} 
\end{subequations}
The spatial trace of the Einstein equation \eqref{spattrace} provides the evolution equation for $K$, and \eqref{spattraceless} is the evolution equation for the traceless part of the extrinsic curvature, which is constrained to vanish in the chosen Newtonian coordinate system.
The trace of the Einstein equations gives the equation of motion \eqref{phieom} for the scalar field $\varphi$. In terms of $\Pi$ (using Eq. \eqref{defPi}) and the metric \eqref{spherMet}, the equation is given by
\begin{equation}
e^{-\Phi}\dot{\Pi}+\Pi^2-K\Pi-e^{2 \Psi}a^{-2}\left(\lapl \varphi+\varphi'^2-\varphi'(\Psi'-\Phi') \right)=e^{-\varphi}\frac{1}{3}\left(2V+\kappa^2 \rho- V_{,\varphi} \right)\,.\label{scalaronequation}
\end{equation}
\end{widetext}
The fluid equation $T^{\mu}{}_{\ \nu;\mu}=0$ determines $\rho$ and $v$, and depends on the metric potenials $\Psi$ and $\Phi$. It splits into energy conservation
\begin{equation}
\partial_t\left(w e^{-3 \Psi} a^3 \rho \right)+  \frac{1}{ar^2}\partial_r\left( r^2 w e^{-3\Psi} a^3 \rho v\right)=0
\end{equation}
and Euler equation
\begin{equation}
\dot{u}_r+\frac{v}{a}u_r'= -e^{\Phi}\left(w \Phi'+\frac{w^2-1}{w} \Psi' \right)\,.
\end{equation}

\section{Linearization}
In the context of cosmological perturbation theory one encounters two types of linearization in the literature. The first \cite{KS82,MFB92} is predicated upon three assumptions; (i) perturbations in the metric are small $\Phi, \Psi \ll 1$, (ii) energy momentum tensor perturbations are small $\delta, v \ll 1$ and (iii) the Einstein and fluid equations are linearized around a background FRW spacetime. This is an excellent approximation in the very early universe, where both fluid and metric perturbations are small. However during structure formation the Newtonian gauge density contrast $\delta$ becomes large
\begin{equation}
\delta\equiv\frac{\rho}{\bar{\rho}}-1 >1\,.
\end{equation}
In spite of this breakdown of assumption (ii) at late times, a miraculously working seemingly inconsistent second linearization scheme is used on subhorizon scales. The metric is assumed to be a Newtonian gauge linearly perturbed FRW metric (i) with $\Phi, \Psi \ll 1$, but (ii) the density contrast $\delta$ is allowed to become non-perturbatively large $\delta>1$, while the velocity $v$ is assumed to remain small $v\ll 1$. The field and fluid (iii) equations are then expanded according to the following scheme, which in case of GR can be expressed solely in terms of the Newtonian metric perturbation \cite{IW06} 
\begin{equation}
\Phi\ll 1\,,\quad \dot{\Phi}\ll \Phi_{,i}/a\,,\quad \Phi_{,i}\Phi_{,j}\ll \Phi_{,ij} \sim \delta \gtrsim 1\,. \label{gradexpan}
\end{equation} 
Applying \eqref{gradexpan} to the Einstein and fluid equations is a procedure known as the quasistatic approximation. It is mixture of expanding in the smallness of $\Phi$ and the smallness of $Ha/k$ on subhorizon scales, the smallness of the velocity $v$ and the smallness of $\dot{\Phi} \lesssim H \Phi$ \cite{R11}. We thus linearize all equations with respect to $\Phi,\dot{\Phi},\varphi,\dot{\varphi},v$ and $\dot{v}$, but not with respect to spatial derivatives of these quantities. 

\subsection{Choice of gauge}
Before we examine the fully non-linear equations to see why this expansion indeed works in the Newtonian gauge, we first consider the scalar part of the linearized comoving synchronous gauge metric \cite{MB95}
\begin{equation}
d s^2= - d t^2 + a^2 \left(\delta_{ij}+A \delta_{ij}+B_{,ij}\right)dx^i dx^j\,. \label{pertmetsync}
\end{equation}
The $0i$ scalar Einstein equation of a pressureless fluid becomes
\begin{equation}
\kappa^2 \bar{\rho} u= \dot{A}\,.
\end{equation}
The scalar velocity perturbation $u$ vanishes in the CDM comoving gauge, such that $\dot{A}=0$. Using the background continuity equation $\bar{ T}^{0\mu}_{\ \ ;\mu}=0$, the perturbed equation $\delta T^{0\mu}_{\ \ ;\mu}=0$ becomes
\begin{equation}
\label{eq:sa1}
\dot{D}= -\frac{1}{2}(3\dot{A}+ \lapl \dot{B})\,,
\end{equation}
with $D\equiv \rho/\bar{\rho}|_{\mathrm{syn}}-1$. The growing mode of the density perturbation evolves as $D\sim a$, and ($\ref{eq:sa1}$) admits the solution
\begin{equation}
A=0\qquad \lapl B=-2 D\,.
\end{equation}
We thus see that the linearized metric written in synchronous gauge \eqref{pertmetsync} becomes nonlinear as soon as the density contrast $\delta\simeq D$ does. This scenario is what one would naively expect, but we will see that a special feature of the Newtonian coordinates is that the expansion \eqref{gradexpan} is consistent.

\subsection{Newtonian gauge discussion}
In the case of standard GR, small velocities $v$ and initially small Newtonian potentials $\Phi$ and $\Psi$, it can be shown using the Lema{\^i}tre-Tolman-Bondi metric  \cite{VanA08} that the solution to the metric-linearized, quasistatic spherically symmetric field equations also solves the fully nonlinear equations. This means that the linearized metric in the Newtonian gauge accurately describes the geometry of spacetime even if density perturbations become large and the spacetime curvature becomes non-linearly distorted. If we accept this, it is clear that the Newtonian coordinates belong to a class which make the approximate FRW symmetry of the metric manifest. One might question whether this property survives if we drop the assumption of spherical symmetry. The answer would appear to be in the affirmative; $\Phi$ and  $\Psi$ are small for all practical purposes in the entire universe (except near Black Holes and Neutron stars). Assuming \eqref{gradexpan}, the Newtonian coordinates have proved very useful without spherical symmetry to model nonlinear structure formation \cite{BS11} and estimating the effect of backreaction of nonlinear structures on the FRW background \cite{GW11}.

It is important to note that a spacetime is not necessarily close to FRW just because the metric is close to FRW. This is because the curvature contains second derivatives of the metric and they are known to become large in the Newtonian gauge $\lapl \Phi \sim \delta$. Two spacetimes can differ significantly despite their metrics being related by a small deformation. Also, even in a spacetime in which the metric is close to FRW and there is negligible backreaction \cite{GW11}, there can still be a large effect on observables, as photons probe both the metric and the curvature \cite{EMR09,R11}.

\subsubsection{General relativity}
We now examine the full non-linear equations \eqref{fofRequations} with $\varphi=0$, and show why the quasistatic expansion is successful. We do this as preparation for the $f(R)$ case, where no exact LTB like solution is known and hence no comparison between the Newtonian linearized and exact solution can be performed\footnote{this comparision was performed in the case of GR in \cite{VanA08}}. Subtracting the background equations
\begin{equation}
3H^2=\bar{\rho}\,,\qquad
 6 \dot{H}+ 9H^2=0\,, \label{Gback}
\end{equation}
\noindent and writing $K=\delta K- 3H$ and $\rho=\bar{\rho}(1+\delta)$, the fully non-linear Einstein equations \eqref{fofRequations} are given by
\begin{widetext}
\begin{subequations}\label{sphEinstpert}
\begin{align}
\frac{1}{3}\delta K^2-H\delta K+2e^{2 \Psi}a^{-2}\left(\lapl \Psi-\frac{1}{2}\Psi'^2\right)&=\kappa^2 \bar{\rho}\left(( w^2-1)+w^2\delta\right)\label{ensphEinstpert}\\
\frac{2}{3}\delta K'&= -\kappa^2\bar{\rho}(1+\delta) w u_r \label{momsphEinstpert}\\
 2 e^{-\Phi}\dot{\delta K}+9(e^{-\Phi}-1)H^2-\delta K^2+6H\delta K-\qquad\qquad\qquad\nonumber\\
 -2 e^{2 \Psi} a^{-2}\left( \lapl(\Psi-\Phi)-\Phi'^2-\frac{1}{2}\Psi'^2+\Phi'\Psi'\right)&=\bar{\rho}(1+\delta)(w^2-1)\label{tracesphEinstpert}\\ \lapl (\Psi-\Phi)-\frac{3}{r}(\Psi'-\Phi')+\Psi'^2-\Phi'^2-2\Psi'\Phi'&=0\label{tracelesphEinstpert}\,.
\end{align}
\end{subequations}
\end{widetext}
\noindent Using the background energy conservation $\bar{\rho} a^3=$const, the fluid equations reduce to
\begin{equation}
\partial_t \left(w e^{-3 \Psi} (1+\delta)\right)+  \frac{1}{ar^2}\left( r^2 w e^{-3\Psi} (1+\delta) v\right)'=0\label{nonlinenerg}
\end{equation}
\begin{equation}
\dot{u}_r+\frac{v}{a}u_r'= -e^{\Phi}\left(w \Phi'+\frac{w^2-1}{w} \Psi' \right)\,.\label{nonlineuler}
\end{equation}
If $\Phi$, $\Psi$ are small initially and we can estimate their spatial derivatives as $a^{-1}\Phi'\sim \Phi/L$, where $L$ is the physical size of the perturbation, then we can infer from \eqref{tracelesphEinstpert} that $\Psi-\Phi=\mathcal{O}(\Phi^2)$.
Inspecting Eq.\,\eqref{tracesphEinstpert} it is then easy to see that if $v$, $\Phi$, $\Psi$ and $\delta K$ are small 
initially, then all source terms for $\dot{\delta K}$ are also small. This prevents 
\begin{equation}
\delta K=-3(e^{-\Phi}-1)H+3e^{-\Phi}\dot{\Psi}
\end{equation}
 and thus also $\Psi$ and $\Phi$ from growing significantly, as long as $v$ remains non-relativistic. Specifically we require $(HL)^2 (1+\delta) v^2\sim \mathcal{O}(\Phi^2)$ and $HL^2\delta K\sim (HL)^3 (1+\delta) v\sim \mathcal{O}(\Phi^2)$ in Eq.\,\eqref{tracesphEinstpert}, where we used \eqref{momsphEinstpert}. Subject to these conditions, the quasistatic approximation will be a consistent expansion and the metric will remain linearly perturbed away from the background, even if $\delta$ becomes nonlinear. 
 We thus arrive at the following equations:
 \begin{subequations}\label{sphEinstlin}
\begin{align}
2 a^{-2}\lapl \Phi&=\kappa^2 \bar{\rho}\delta \label{ensphEinstlin}\\
\dot{\delta} + \frac{1}{ar^2}\partial_r\left( r^2(1+\delta) v\right)&=0\label{linenergycon}\\
\dot{v}+v H+\frac{v}{a}v'&= -\frac{1}{a}\Phi'\,.\label{lineuler}
\end{align}
\end{subequations}
\noindent They are sufficient to determine $\Phi$, $\delta$ and $v$ from initial data. At the initial time we have to ensure that the constraint  \eqref{momsphEinstpert}
\begin{equation}
\frac{2}{3}\delta K'= 2 \Phi' H+2 \dot{\Phi}'= -\kappa^2\bar{\rho}(1+\delta) a v \label{momconstrlin}
\end{equation}
is fulfilled, such that the subsequent evolution of $\delta K$ and thus \eqref{tracesphEinstpert} is irrelevant. Note that quasistatic approximation and metric linearization break down near the final stages of the collapse, where a Black Hole will form. We stop our simulations well before this time, making the reasonable assumption that the duration of the final stage of collapse is negligible.
\subsubsection{f(R)}
We now apply a similar argument to the fully nonlinear $f(R)$ Einstein equations \eqref{fofRequations}. Again subtracting the background equations
\begin{subequations} \label{fofRbackground}
\begin{align}
3 H(H+\bar{\Pi})&=e^{-\bar{\varphi}}(\kappa^2 \bar{\rho}+\frac{1}{2}\bar{V})\\
2 \dot{H}+3H^2+\dot{\bar{\Pi}}+ \bar{\Pi}^2+2H\bar{\Pi}  &=\frac{1}{2}e^{-\bar{\varphi}}\bar{V}\\
\dot{\bar{\Pi}}+\bar{\Pi}^2+3H\bar{\Pi}&=e^{-\bar{\varphi}}\frac{1}{3}\left(2\bar{V}+\kappa^2 \bar{\rho}- \bar{V}_{,\varphi} \right)\,,
\end{align}
\end{subequations}
\noindent and defining $\delta \varphi\equiv \varphi-\bar{\varphi}$, $\delta\!V\equiv V-\bar{V}$ and $\delta\!V_{,\varphi}\equiv V_{,\varphi}-\bar{V}_{,\varphi}$, the modified Einstein equations \eqref{fofRequations} take the following form.
\begin{widetext}
Energy constraint
\begin{subequations}\label{fofRequationspert}
\begin{align}
\frac{1}{3}\delta K^2-H\delta K-\delta K \bar\Pi+3H\delta \Pi-\delta K\delta \Pi +\qquad\qquad\qquad\nonumber\\ +e^{2 \Psi}a^{-2}\left(\lapl (2\Psi-\varphi)-\Psi'^2+\Psi'\varphi'-\varphi'^2\right)&= e^{-\varphi}\left(\kappa^2 \bar{\rho}\left((w^2-1)+ w^2\delta\right)+\frac{1}{2}\delta\!V\right)\nonumber \\
&\qquad -e^{-\bar{\varphi}}\left(\kappa^2 \bar{\rho}+\frac{1}{2}\bar{V} \right)(1-e^{-\delta\varphi})\,,\label{fofRconstraintpert}
\end{align}
momentum constraint
\begin{equation}
\label{eq:t1}
 \frac{2}{3}\delta K'-\left(\delta \Pi'+(\bar{\Pi}+\delta{\Pi}) \varphi'+\frac{1}{3}\delta K \varphi'-H \varphi' \right) = -e^{-\varphi} \kappa^2\bar{\rho}(1+\delta) w u_r\,,
 \end{equation}
evolution equation
 \begin{multline}
 e^{-\Phi} (2 \dot{\delta K}-3\dot{\delta\Pi})+3(1-e^{-\Phi})(2\dot{H}+\dot{\bar{\Pi}})+6H\delta K-\delta K^2+2\delta K \bar{\Pi}-\qquad\\-6H\delta \Pi+2\delta K \delta \Pi-6\bar{\Pi}\delta \Pi-3\delta \Pi^2-2e^{2 \Psi} a^{-2} \lapl(\Psi-\Phi-\varphi)-\qquad\qquad\\-2e^{2 \Psi} a^{-2}\left( -\Phi'^2-\frac{1}{2}\Psi'^2+\Phi'\Psi'-\varphi'^2+\Psi' \varphi' -\frac{3}{2}\Phi' \varphi' \right)\\= (1-e^{-\delta \varphi})e^{-\bar{\varphi}}\frac{3}{2}\bar{V}-e^{-\varphi}\left(\kappa^2\rho (1-w^2)+\frac{3}{2}\delta\!V\right)\,,\label{fofRspattracepert}
 \end{multline}
Newtonian gauge condition
 \begin{equation}
 \lapl (\Psi-\Phi-\varphi)-\frac{3}{r}(\Psi'-\Phi'-\varphi')+\Psi'^2-\Phi'^2-\varphi'^2-2\Psi'\Phi'-2\Psi' \varphi'= 0\,.\label{fofRspattracelepert}
 \end{equation} 
and scalar field equation
\begin{multline}
e^{-\Phi}\dot{\delta\Pi}+\delta \Pi^2-\delta K \delta \Pi+2\delta \Pi \bar{\Pi}-\delta K \bar{\Pi}+3H\delta\Pi-e^{2 \Psi}a^{-2}\left(\lapl \varphi+\varphi'^2-\varphi'(\Psi'-\Phi') \right)\\=(e^{-\delta \varphi}-1)e^{-\bar{\varphi}}\frac{1}{3}\left(2\bar{V}+\kappa^2 \bar{\rho}- \bar{V}_{,\varphi} \right)+e^{-\varphi}\frac{1}{3}(2\delta\!V+\kappa^2 \bar{\rho}\delta-\delta\!V_{,\varphi})\,.\label{scalaronpert}
\end{multline}
\end{subequations}
\end{widetext}
The argument for using the quasistatic approximation in the $f(R)$ field equations contains additional caveats. The reason is the following. Some of the source terms for $\dot{\delta \Pi}$ in \eqref{scalaronpert} are known to become large during the evolution, namely $\lapl \varphi$, $\delta$ and $\delta\!V_{,\varphi}$, hence it is not a priori clear that $\dot{\delta \Pi}$ will remain small (and similarly $\dot{\delta K}$). However if $\delta \varphi$, $\Psi$ and $\Phi$ are small initially, and relations such as $a^{-1}\Phi'\sim \Phi/L$ hold for all three variables individually, then from \eqref{fofRspattracelepert} we can deduce that $\Psi-\Phi-\varphi\sim \mathcal{O}(\Phi^2)$. If in addition $v$, $\delta\!V$, $\delta\Pi$ and $\delta K$ are small then the evolution equation \eqref{fofRspattracepert} tells us that the combination $2 \dot{\delta K}-3\dot{\delta\Pi}$ will also be small. However we cannot infer that $\delta\Pi$ and $\delta K$ both remain small individually, so we can make no definite statement regarding the magnitudes of $\delta \varphi$, $\Psi$ and $\Phi$. We can argue that it is plausible that if the combination $2 \dot{\delta K}-3\dot{\delta\Pi}$ is small then $\dot{\delta K}$ and $\dot{\delta\Pi}$ are small separately, unless there is some form of cancelation. With this extra assumption we can write down the relevant equations. Subject to $v, \delta\!V \ll 1$, \eqref{fofRconstraintpert} and \eqref{scalaronpert} become
\begin{subequations}\label{fofRequationslin}
\begin{align}
a^{-2}\lapl (2\Phi+\varphi)&= e^{-\bar{\varphi}}\kappa^2 \bar{\rho}\delta\,, \\
-a^{-2}\lapl \varphi&=e^{-\bar{\varphi}}\frac{1}{3}(\kappa^2 \bar{\rho}\delta-\delta\!V_{,\varphi})\,.
\end{align}
\end{subequations}
\noindent and again we need to enforce the constraint equation
\begin{align}
 \frac{2}{3}\delta K'-\left(\delta \Pi'+(\bar{\Pi}+\delta{\Pi}) \varphi'+\frac{1}{3}\delta K \varphi'-H \varphi' \right)\nonumber\\ = -e^{-\varphi} \kappa^2\bar{\rho}(1+\delta) a v\,,\label{fofRmomconstrlin}
\end{align}
on the initial time slice.

The $f(R)$ model that we are using \eqref{powerlaw} allows for further simplifications. The background value of the scalar field today is of order $\bar{\varphi}\simeq-\epsilon$. This makes $\bar{\varphi}$ itself quasistatic for $\epsilon \ll 1$ and we can neglect $\bar{\Pi}$, $\dot{\bar{\Pi}}$ compared to $H$ and $\dot{H}$ respectively in \eqref{fofRbackground}. Using $V\simeq R\varphi -f \simeq 2\Lambda$, the relevant background equations reduce to $\Lambdaup$CDM
\begin{subequations} \label{fofRbackgroundrel}
\begin{align}
3 H^2&=\kappa^2 \bar{\rho}+\Lambda\\
2 \dot{H}+3H^2&=\Lambda\,.
\end{align}
\end{subequations}
 If we take initial conditions during matter domination, all $f(R)$ terms in \eqref{fofRmomconstrlin} are negligible such that we can use the GR momentum constraint \eqref{momconstrlin}. In addition, since $\dot{\Phi}=0$ during the matter era, the initial velocity $v$ must satisfy the GR condition
\begin{equation}
2 \Phi' H= -\kappa^2\bar{\rho}(1+\delta) a v \label{fofRmomconstrlinfinal}
\end{equation}
Eqs.\,\eqref{fofRequationslin} together with the fluid equations \eqref{linenergycon} and \eqref{lineuler} completely determine the nonlinear spherical collapse provided all quantities are initially perturbative and subject to the conditions; (i) $\delta\!V$ is comparably small to $v^2(1+\delta)$ and (ii) $v\sim HL$. This involves the extra assumption that $2 \delta K -3\delta\Pi\simeq 6\dot{\Psi}-3\dot{\varphi}$ has the same order of magnitude as both $\delta K\simeq 6\dot{\Psi}$ and $\delta\Pi\simeq \dot{\varphi}$.
So the full set of relevant equations is

\begin{subequations}\label{fofRequationslinfinal}
\begin{align}
a^{-2}\lapl \Phi&= \frac{2}{3}\kappa^2 \bar{\rho}\delta-\frac{1}{6} \delta\!V_{,\varphi}\,, \label{poieq}\\
a^{-2}\lapl \varphi&=\frac{1}{3}(\delta\!V_{,\varphi}-\kappa^2 \bar{\rho}\delta)\label{phieq}\\
\dot{\delta} + \frac{1}{ar^2}\partial_r\left( r^2(1+\delta) v\right)&=0\\
\dot{v}+v H+\frac{v}{a}v'&= -\frac{1}{a}\Phi'\,.\label{eulereq}
\end{align}
\end{subequations}
Note that one could replace equation \eqref{poieq} or \eqref{phieq} by
\begin{equation}
(2 \Phi+\varphi)' H+ (2 \dot{\Phi} +\dot{\varphi})'= -\kappa^2\bar{\rho}(1+\delta) a v \label{fofRmomconstrlinfinal2}
\end{equation}
which is the simplified \eqref{fofRmomconstrlin} valid in the quasistatic approximation. One advantage of \eqref{fofRmomconstrlinfinal2} compared to \eqref{phieq} is that only the former is linear in $\varphi$.

The assumptions made above regarding the size of $\dot{\Phi}$ and $\dot{\varphi}$ may be unjustified in situations where the effective potential of $\varphi$ suddenly changes, such as during the onset of the chameleon mechanism (see Fig.\,\ref{fig:chameleonshell}). Another example might be the oscillation of  $\varphi$ during the emission of monopole radiation. In these situations $\Phi$ may compensate the time dependence of $\varphi$ such that the combination $2 \Phi +\varphi$ might remain quasi-static, but it is not clear that any of the assumptions made will continue to hold. Since we cannot say anything definite about the validity of equations \eqref{fofRequationslinfinal}, we check during the numerical solution of the equations that all neglected terms stay much smaller than the terms appearing in \eqref{fofRequationslinfinal}, and also that the neglected equation \eqref{fofRspattracelepert} is satisfied. 

While performing these checks we noticed that $\dot{\Phi}\ll\Phi '/a$ is never satisfied well within and far outside the density perturbation; rather we find $\dot{\Phi}\gg\Phi '/a$ in these regions. This is not an $f(R)$ artifact, but is simply a consequence of the boundary conditions at $r=0$ and $r=\infty$, where all spatial derivatives approach zero. It seems that $\dot{\Phi}\ll\Phi '/a$ is not globally required to ensure a quasistatic evolution.

\section{peaks theory shape function}
\label{section:peaks}
Consider a gaussian random field $\delta(z_i,\mathbf{x},R)$ smoothed with a window function $W(kR)$ over the comoving scale $R$. The properties of this field are completely determined by its two-point correlation function $\xi(r=|\mathbf{x}-\mathbf{y}|,R)$, or equivalently, its power spectrum $P(k,R)=W(k R)^2 P(k)$. The mean shape 
\begin{equation}
\delta_i(r,R)=\langle \delta(z_i,\mathbf{x},R)| \mathrm{peak},\nu\rangle
\end{equation}
around a peak of height $\nu=\delta_{i,0}/\sigma(z_i,R)$ can be expressed in terms of the autocorrelation function $\xi(r,R)$ and its first, second and forth derivative with respect to $r$. Following Appendix A-D and Section VII of \cite{BBKS86} we arrive at\footnote{Note that eq.\,(D6) and (7.10) of \cite{BBKS86} do in fact coincide with each other and with \eqref{peaksformula} after rescaling the radial coordinate. We thank Ravi Sheth for pointing this out to us. Eq.\,\eqref{peaksformula} was be obtained by averaging over $x,y$ and $z$ in eq. (D3) of \cite{BBKS86}.}
\begin{equation}
\delta_i(r,R)= \frac{\sigma_0 \nu}{1-\gamma^2}\left(\psi+\gamma \frac{\sigma_0}{\sigma_2}\lapl \psi- \frac{\langle x |  \mathrm{peak},\nu\rangle}{\nu} \left(\gamma \psi+\frac{\sigma_0}{\sigma_2} \lapl \psi \right) \right)\label{peaksformula}
\end{equation}
with variance $\sigma\equiv \sigma_0$ and the first two moments $\sigma_1,\sigma_2$ are given by

\begin{equation}
\sigma_i= \int_0^\infty \frac{k^2 d k}{2\pi^2} P(k,R) k^{2i}\,,
\end{equation}
and with $\psi\equiv\xi/\sigma_0^2$,  $x\equiv - \lapl \delta|_{\mathbf{x}=0}/\sigma_2$ and $\gamma\equiv \sigma_1^2/(\sigma_0 \sigma_2)$. The mean central curvature $x$  of a peak is approximately given by eq. (6.13) and (6.14) of \cite{BBKS86}
\begin{align}
\langle x |  \mathrm{peak},\nu\rangle &= \gamma \nu + \theta\\
\theta&=\frac{3(1-\gamma^2)+(1.216-0.9 \gamma^4)\exp[-\gamma/2\,(\gamma\nu/2)^2]}{[3(1-\gamma^2)+0.45+(\gamma\nu/2)^2]^{1/2}+\gamma \nu/2}\,.
\end{align}
With a gaussian filter $W=\exp(-k^2 R^2/2)$ and primordial spectrum $P_0(k)\sim k^{n_s}$, the mean shape $\delta_i(r,R)$ is given by
\begin{widetext}
\begin{multline}
\delta_i(r,R)=T(r)*\delta_{i,0}\left\{
  \,  _1F_1\left(\frac{n_s+3}{2};\frac{3}{2};-\frac{r^2}{4 R^2}\right)\phantom{\frac{r^2 e^{-\frac{1}{8} \left(\frac{n_s+3}{n_s+5}\right)^{3/2} \nu ^2} \left(((-n_s-8)
   n_s-15) e^{\frac{1}{8} \left(\frac{n_s+3}{n_s+5}\right)^{3/2} \nu ^2}+n_s ((-0.05267
   n_s-1.28467) n_s-7.0967)-11.15\right) \, _1F_1\left(\frac{n_s+5}{2};\frac{5}{2};-\frac{r^2}{4
   R^2}\right)}{\nu\sqrt{(n_s+3) (n_s+5)} (n_s+5) R^2 \left(2 \sqrt{\frac{(0.25 n_s+0.75)
   \nu ^2+0.45 n_s+8.25}{n_s+5}}+\sqrt{\frac{n_s+3}{n_s+5}} \nu \right)}}\right.+\\   \left. \frac{r^2 e^{-\frac{1}{8} \left(\frac{n_s+3}{n_s+5}\right)^{3/2} \nu ^2} \left(((-n_s-8)
   n_s-15) e^{\frac{1}{8} \left(\frac{n_s+3}{n_s+5}\right)^{3/2} \nu ^2}+n_s ((-0.05267
   n_s-1.28467) n_s-7.0967)-11.15\right) \, _1F_1\left(\frac{n_s+5}{2};\frac{5}{2};-\frac{r^2}{4
   R^2}\right)}{\nu\sqrt{(n_s+3) (n_s+5)} (n_s+5) R^2 \left(2 \sqrt{\frac{(0.25 n_s+0.75)
   \nu ^2+0.45 n_s+8.25}{n_s+5}}+\sqrt{\frac{n_s+3}{n_s+5}} \nu \right)}\right\}\,,\label{peaksshape}
\end{multline}
\end{widetext}
where $_1F_1$ is the confluent hypergeometric function and $T(r)$ the post-recombination transfer function in the Newtonian gauge, related to the synchronous gauge function $T_\mathrm{sync}(k)$ via $T(k)=(1+3 a^2 H^2/k^2)T_\mathrm{sync}(k)$. Here $T(k)$ is normalized as $T(k\!\!=\!\!0)=1$. Rather than applying the actual power spectrum $P_i\sim k^{n_s}T(k)^2$ at redshift $z_i=200$  to the peaks theory shape formula \eqref{peaksformula}, we use the primordial powerspectrum $P_0\sim k^{n_s}$ to calculate the mean shape at very early times on superhorizon scales and use the transfer function $T(k)$ to evolve this shape to subhorizon scales after matter radiation equality.
Note that the primordial amplitude of $P(k)$ and its linear growth is irrelevant here, as the $k$- and $a$-dependence factorize in linear perturbation theory and we are free to choose our initial $\nu$ at $z_i$. 
The primordial shape function in $k$-space $\delta_0(k)=\int_0^\infty d r r^2 \frac{\sin k r}{k r} \delta_0(r)$ is given by
\begin{widetext}
\begin{multline}
\delta_0(k,R)=\delta_{i,0}\frac{1}{4} \pi  (n_s+5) R^3 e^{-k^2 R^2} (k R)^{n_s}\cdot  \\ \cdot \left(\frac{\sqrt{\frac{n_s+3}{n_s+5}}}{ \nu \Gamma
   \left(\frac{n_s+5}{2}\right)}
   \left(\frac{e^{-\frac{1}{8} \left(\frac{n_s+3}{n_s+5}\right)^{3/2} \nu ^2} \left((12 n_s+60)
   e^{\frac{1}{8} \left(\frac{n_s+3}{n_s+5}\right)^{3/2} \nu ^2}+(0.632 n_s+13.52)
   n_s+44.6\right)}{(n_s+5)^2 \left(2 \sqrt{\frac{(0.25 n_s+0.75) \nu ^2+0.45
   n_s+8.25}{n_s+5}}+\sqrt{\frac{n_s+3}{n_s+5}} \nu
   \right)}+\sqrt{\frac{n_s+3}{n_s+5}} \nu \right) \left(2 k^2
   R^2-n_s-3\right)\right. \\\left. \phantom{\frac{\sqrt{\frac{n_s+3}{n_s+5}}}{ \Gamma
   \left(\frac{n_s+5}{2}\right)}}+\frac{(n_s+3)  \left(-2 k^2 R^2+n_s+3\right)}{2 \Gamma
   \left(\frac{n_s+7}{2}\right)}+\frac{4 }{(n_s+5) \Gamma \left(\frac{n_s+3}{2}\right)}\right) \label{peaksshapeprim}
   \end{multline}
\end{widetext}
and one obtains \eqref{peaksshape} via the integral

$$\delta_i(r,R)=\frac{2}{\pi} \int_0^\infty d k k^2 \delta_0(k,R) \frac{\sin k r}{k r} T(k)\,.$$ 
\bibliography{collapse}
\end{document}